\documentclass[a4]{article}
\usepackage[T1]{fontenc}
\usepackage[latin1]{inputenc}
\usepackage{amsmath, amssymb, amsthm}
\usepackage[english]{babel}
\usepackage{indentfirst}
\usepackage{booktabs}
\usepackage{array}
\usepackage{caption,appendix}
\usepackage{geometry}
\usepackage{float}
\usepackage{cancel}
\usepackage{bbold}
\usepackage{authblk}
\usepackage{cite}
\usepackage{xcolor}
\allowdisplaybreaks
\numberwithin{equation}{section}

\theoremstyle{definition}

\begin{document}
\author[1,2,3]{Salvatore Capozziello \thanks{capozziello@unina.it}}
\author[4]{Maurizio Capriolo \thanks{mcapriolo@unisa.it} }
\author[4,5]{Gaetano Lambiase \thanks{glambiase@unisa.it}}
\affil[1]{\emph{Dipartimento di Fisica "E. Pancini", Universit\`a di Napoli {}``Federico II'', Compl. Univ. di
		   Monte S. Angelo, Edificio G, Via Cinthia, I-80126, Napoli, Italy, }}
\affil[2]{\emph{INFN Sezione  di Napoli, Compl. Univ. di
		   Monte S. Angelo, Edificio G, Via Cinthia, I-80126,  Napoli, Italy,}}
\affil[3]{\emph{Scuola Superiore Meridionale, Largo S. Marcellino 10,  I-80138,  Napoli, Italy,}}
\affil[4]{\emph{Dipartimento di Fisica Universit\`a di Salerno, via Giovanni Paolo II, 132, Fisciano, SA I-84084, Italy.} }
\affil[5]{\emph{INFN Sezione  di Napoli,Gruppo Collegato di Salerno, via Giovanni Paolo II, 132, Fisciano, SA I-84084, Italy.} }
\date{\today}
\title{\textbf{Gravitational tensor and  scalar modes  in $f(Q,B)$ non-metric gravity}}
\maketitle

\begin{abstract} 
We investigate gravitational waves  generated in $f(Q,B)$ non-metric gravity, i.e.,  a theory of  gravity described by a non-metric compatible connection, free of torsion and curvature. It is an extension of  symmetric teleparallel gravity, equipped with a boundary term $B$.  This theory exhibits gravitational waves regardless of the gauge adopted: they are the standard massless tensors  plus  a massive scalar gravitational wave like in the case of $f(R)$ gravity. It is precisely the boundary term $B$ that generates the massive scalar mode with an effective mass $m_{B}$ associated to a Klein-Gordon equation in the linearized boundary term.  As in $f(Q)$  gravity also in $f(Q,B)$ non-metric gravity, a free test particle follows a geodesic motion due to the covariant conservation with respect to the Levi-Civita connection of the energy and momentum densities on-shell. Therefore, in $f(Q,B)$ gravity, the proper acceleration between two neighboring worldlines traveled by two free point-like particle is governed by a first-order geodesic deviation equation in the metric perturbation $h_{\mu\nu}$. Thanks to this approximate linear equation, $f(Q,B)$ non-metric gravity shows three polarization modes: two massless transverse tensor radiation modes, with helicity equal to 2, reproducing the standard plus and cross modes, exactly as in General Relativity, and an additional massive scalar wave mode with transverse polarization of zero helicity.  We obtain the same result both by considering the coincidence gauge and by leaving the gauge free. This happens because, at first order, the boundary term $B$ remains unchanged as well as the linearized field equations of  metric tensor  and connection in vacuum.  Furthermore, we  derive the energy and momentum balance equations and the equations of motion of the deviation, both in a generic theory of gravitation, where the matter stress-energy tensor $T^{\alpha}_{\phantom{\alpha}\beta}$ is Levi-Civita covariantly not conserved, i.e., $\mathcal{D}_{\alpha}T^{\alpha}_{\phantom{\alpha}\beta}\neq 0$. In summary, three degrees of freedom propagate in the $f(Q,B)$ linearized theory with amplitudes $\tilde{h}^{(+)}$ and $\tilde{h}^{(\times)}$ for tensor modes and amplitude $\tilde{h}^{(s)}$ for the scalar mode. Specifically, both $f(Q,B)$ and  $f(R)$  gravity  involve the same massive transverse scalar perturbation.  
\end{abstract}

\section{Introduction}
General Relativity (GR)  passed  excellent probes  such as the Solar System tests in the weak field limit,  the predicted  existence of gravitational waves (GWs)  and black holes, and the  compatibility with the cosmological model based on the cosmological principle and expansion as well as the Big Bang theory.  However, to explain the behavior of the Universe in the early and late times, dynamics of galaxies, and large scale structure researchers  have been forced to hypothesize the existence of unknown cosmic components as dark energy and dark matter, in order to preserve the validity of General Relativity.  To date, there is no direct experimental evidence for the existence of these two exotic components at fundamental level,  and more and more physicists are proposing to modify  geometry rather than the matter sector in order to fit dynamics~\cite{Starobinsky, Nojiri1, Caprep, Nojiri2, Clifton, DeFelice, Faraoni, Odi2,Cai}.  We can also add that  GR fails in strong gravity regime and that it is a non-renormalizable theory, an indispensable condition for its quantization.  In summary, GR presents problem both at IR and UV scales. As a consequence, theories of gravity based on other geometric invariants, over the standard Ricci scalar $R$, are going to be proposed.  Among these  new proposal, 
metric-affine theories are assuming a relevant role to cure GR shortcomings. In this enlarged perspective,   curvature,  torsion and  non-metricity can be dealt under the same standard to represent gravitational dynamics.

When we turn off both the curvature and the torsion in the connection, we get the so called symmetric teleparallel gravity (STG), a class of theories  based on manifold endowed with a flat and torsionless connection $\Gamma$ and a metric tensor $g$, where the gravitational interaction is completely described by the non-metricity scalar $Q$.  In  the Palatini formalism, these two geometric objects, metric tensor and STG connection, are independent and obey related field equations.  

In the {\it coincidence gauge}, where we set the STG connection to zero, the theory of gravity described by the gravitational action, linear in the non-metricity scalar $Q$,  is equivalent to GR up to boundary terms, namely  the well-known Symmetric Teleparallel Equivalent of General Relativity (STEGR)~\cite{BJHK,JHK}. Similarly, the theory of gravity described by the linear gravitational action in the torsion scalar $T$, expressed in a tetrad basis and satisfying the tetrad postulate,  is equivalent to GR, namely the so-called Teleparallel Equivalent of General Relativity (TEGR)~\cite{Aldrovandi}. 

Nonlinear extensions of  TEGR and STEGR can be obtained by  analytic functions $f$ as $f(T)$~\cite{Cai} and $f(Q)$~\cite{LaHe} gravity, as well as from the linear Einstein-Hilbert action in the curvature scalar $\mathcal{R}$, we can move on  $f(R)$ gravity~\cite{Caprep,Nojiri2,Clifton, DeFelice,Faraoni}.  If we do not neglect the boundary terms $B$, we can extend them further and thus obtain $f(T,B)$ and $f(Q,B)$ gravity, in teleparallel and symmetric teleparallel framework, respectively.  

The $f(Q)$ gravity is enjoying great success both in astrophysics and cosmology~\cite{App1,App2,App3,App4,App5,App6,App7,JHKP,Mehdi,BVC,Nojiri:2024hau,NO_1,DPRC } as well as in black hole and wormhole solutions.   The analysis of gravitational waves, generated in $f(Q)$ gravity, is noteworthy as well as the study of  propagating degrees of freedom (DOFs). In coincident gauge   $f(Q)$ gravity there are up to eight DOFs, but the ghost scalar mode cannot propagate~\cite{HYKNQ,Hu:2022anq,DAmbrosio:2023asf,GJCK}.  This STG model, based  only on the non-metricity $Q$ scalar, produces only massless transverse tensor GWs but no scalar mode, exactly as the two plus and cross polarization modes typical of GR, both in the coincident gauge, see~\cite{HPUS, SFSGS,CCN2,Sebastian,Jose,XLHL,NF,CONKOI,BJHK} and in any gauge~\cite{CC}.  The production of gravitational waves in a gravity model can be addressed only after obtaining the related gravitational energy-momentum pseudo-tensor, as done in non-local gravity~\cite{CCL1}, in higher-order Riemannian gravity~\cite{CCL, CCT, ACCA}, and in teleparallel gravity~\cite{Capozziello:2018qcp}. The propagation of gravitational waves in non-local gravity has been analyzed in  ~\cite{CAPRIOLOM, CCCQG2021, CCN}, while the propagation of gravitational waves in teleparallel and Riemannian gravity of any order with their constraints has been developed in ~\cite{CCC, CCC1, CCC2, Lambiase:2020vul}.  

In this paper, we are going to extend the study of  GWs to symmetric teleparallel non-metric gravity retaining the boundary term $B$,  i.e.  $f(Q,B)$ gravity. Here, we  analyze whether the $B$ term can induce a further massive scalar mode beyond the two standard tensor modes of GR and $f(Q)$.  We will adopt both the coincident gauge and a free gauge.  Firstly we will derive both energy and momentum balance equations and deviation equation in a generic gravitational theory with $\mathcal{D}_{\alpha}T^{\alpha}_{\phantom{\alpha}\beta}\neq 0$ and then we will examine whether, in $f(Q,B)$ gravity, the divergence $\mathcal{D}_{\alpha}T^{\alpha}_{\phantom{\alpha}\beta}$ vanishes, in order to establish the worldlines traveled by  test bodies and hence the exact deviation equation.  Thus, we will find the metric and connection field equations in coincident and free gauge.  Finally,   polarizations and helicities of GW tensor and scalar modes, solutions of linearized field equations in vacuum, will be analyzed using the deviation equation expressed in a particular locally inertial reference system. 

The paper is organized as follows. In Section~\ref{A}, we briefly review the basic geometric concepts of the  STG. In Section~\ref{B}, we write the  two useful formulas of  energy and momentum balance in generic theories of gravity where the energy-momentum tensor of matter is covariantly not conserved, if the covariant derivative is defined by the Levi-Civita  (LC) connection.  Under the same conditions of non-vanishing  LC covariant divergence of matter tensor, we obtain the  deviation equation in Section~\ref{C}.  Then,  in Section~\ref{D} and its subsections, we discuss the main features of  $f(Q,B)$ non-metric gravity.  More precisely, in Subsection~\ref{DA}, we derive the field and connection equations from the variational principle. In Subsection~\ref{DB}, we prove that the matter energy-momentum tensor is LC covariant divergence-free and so we obtain the motion equation of a particle and the deviation equation in Subsection~\ref{DC}. Section~\ref{DD} is devoted to the  linearization of  field and connection equations in coincident gauge and in any gauge, respectively.  Finally we look for gravitational wave solutions  in Section~\ref{DF}, while, in Section~\ref{DG}, through the deviation equation, we analyze the possible massive and massless tensor and scalar modes of GWs.  The main result is that, like in metric $f(R)$ gravity, we obtain the two GR massless modes plus a further scalar mode,  due to the presence of the boundary term $B$ which behaves as an additional scalar field. Section \ref{E} is devoted to discussion and conclusions. Several useful formulas are reported in the Appendix A.

\section{Symmetric teleparallel gravity: The geometric setting}\label{A}
In  metric-affine gravity, we can uniquely decompose a general affine connection $\Gamma^{\alpha}_{\phantom{\alpha}\mu\nu}$, equipped with non-metricity, torsion and curvature, as~\cite{RV,JHK,CDF,DKC,JHK1}
\begin{equation}\label{1}
\Gamma^{\alpha}_{\phantom{\alpha}\mu\nu}=\genfrac\{\}{0pt}{1}{\alpha}{\mu\nu}+K^{\alpha}_{\phantom{\alpha}\mu\nu}+L^{\alpha}_{\phantom{\alpha}\mu\nu}\ ,
\end{equation}
where $\genfrac\{\}{0pt}{1}{\alpha}{\mu\nu}$ are the Christoffel symbols defined as
\begin{equation}\label{2}
\genfrac\{\}{0pt}{1}{\alpha}{\mu\nu}=\frac{1}{2}g^{\alpha\lambda}\left(\partial_{\mu}g_{\lambda\nu}+\partial_{\nu}g_{\lambda\mu}-\partial_{\lambda}g_{\mu\nu}\right)\ ,
\end{equation}
$K^{\alpha}_{\phantom{\alpha}\mu\nu}$ is the contortion tensor defined from the torsion tensor $T^{\alpha}_{\phantom{\alpha}\mu\nu}$ 
\begin{equation}\label{3}
T^{\alpha}_{\phantom{\alpha}\mu\nu}=2\Gamma^{\alpha}_{\phantom{\alpha}[\mu\nu]}\ ,
\end{equation}
as
\begin{align}\label{4}
K^{\alpha}_{\phantom{\alpha}\mu\nu}&=\frac{1}{2}g^{\alpha\lambda}\left(T_{\mu\lambda\nu}+T_{\nu\lambda\mu}+T_{\lambda\mu\nu}\right)\\
&=\frac{1}{2}T^{\alpha}_{\phantom{\alpha}\mu\nu}+T_{(\mu\phantom{\alpha}\nu)}^{\phantom{(\mu}\alpha}\ ,
\end{align}
antisymmetric in the following two indices 
\begin{equation}\label{5}
K_{\alpha\mu\nu}=-K_{\nu\mu\alpha}\ ,
\end{equation}
whose the antisymmetric part is given by
\begin{equation}\label{6}
K^{\alpha}_{\phantom{\alpha}[\mu\nu]}=\frac{1}{2}T^{\alpha}_{\phantom{\alpha}\mu\nu}\ ,
\end{equation}
and $L^{\alpha}_{\phantom{\alpha}\mu\nu}$ is the disformation tensor, symmetric in second and third index  
\begin{align}\label{7}
L^{\alpha}_{\phantom{\alpha}\mu\nu}&=-\frac{1}{2}g^{\alpha\lambda}\left(Q_{\mu\lambda\nu}+Q_{\nu\lambda\mu}-Q_{\lambda\mu\nu}\right)\ ,\\
&=\frac{1}{2}Q^{\alpha}_{\phantom{\alpha}\mu\nu}-Q_{(\mu\phantom{\alpha}\nu)}^{\phantom{(\mu}\alpha}\ ,
\end{align}
\begin{equation}\label{8}
L^{\alpha}_{\phantom{\alpha}[\mu\nu]}=0\ ,
\end{equation}
where $Q_{\alpha\mu\nu}$ is the non-metricity tensor, symmetric in the second and third index, defined as 
\begin{equation}\label{9}
Q_{\alpha\mu\nu}=\nabla_{\alpha}g_{\mu\nu}=\partial_{\alpha}g_{\mu\nu}-\Gamma^{\beta}_{\phantom{\beta}\alpha\mu}g_{\beta\nu}-\Gamma^{\beta}_{\phantom{\beta}\alpha\nu}g_{\beta\mu}\ ,
\end{equation}
\begin{equation}\label{10}
Q_{\alpha[\mu\nu]}=0\ .
\end{equation}
STG is a particular non-metric theory of gravity where the curvature and the torsion of affine connection vanish, that is,
\begin{equation}\label{11}
R^{\lambda}_{\phantom{\lambda}\mu\nu\sigma}=\Gamma^{\lambda}_{\phantom{\lambda}\mu\sigma,\nu}-\Gamma^{\lambda}_{\phantom{\lambda}\mu\nu,\sigma}+\Gamma^{\beta}_{\phantom{\beta}\mu\sigma}\Gamma^{\lambda}_{\phantom{\lambda}\beta\nu}-\Gamma^{\beta}_{\phantom{\beta}\mu\nu}\Gamma^{\lambda}_{\phantom{\lambda}\beta\sigma}=0\ ,
\end{equation}
and
\begin{equation}\label{12}
T^{\alpha}_{\phantom{\alpha}\mu\nu}=0\ .
\end{equation}
So, the STG connection~\eqref{1} can be rewritten as 
\begin{equation}
\Gamma^{\alpha}_{\phantom{\alpha}\mu\nu}=\genfrac\{\}{0pt}{1}{\alpha}{\mu\nu}+L^{\alpha}_{\phantom{\alpha}\mu\nu}\ .
\end{equation}
Specifically in GR, where the connection is symmetric and metric compatible,  Eq.~\eqref{1} becomes 
\begin{equation}
\Gamma^{\alpha}_{\phantom{\alpha}\mu\nu}=\genfrac\{\}{0pt}{1}{\alpha}{\mu\nu}\ ,
\end{equation}
that is, the STG connection $\Gamma^{\alpha}_{\phantom{\alpha}\mu\nu}$ can be reduced to the LC connection that we rewrite as $\hat{\Gamma}^{\alpha}_{\phantom{\alpha}\mu\nu}$.
The non-metricity scalar is defined as
\begin{align}\label{14}
Q\equiv&-\frac{1}{4}Q_{\alpha\beta\gamma}Q^{\alpha\beta\gamma}+\frac{1}{2}Q_{\alpha\beta\gamma}Q^{\gamma\beta\alpha}+\frac{1}{4}Q_{\alpha}Q^{\alpha}-\frac{1}{2}Q_{\alpha}\widetilde{Q}^{\alpha}\\
=&\, g^{\mu\nu}\Bigl(L^{\alpha}_{\phantom{\alpha}\beta\mu}L^{\beta}_{\phantom{\beta}\nu\alpha}-L^{\alpha}_{\phantom{\alpha}\beta\alpha}L^{\beta}_{\phantom{\beta}\mu\nu}\Bigr)\ ,
\end{align}
while  the two traces of non-metricity tensor are
\begin{equation}\label{15}
Q_{\alpha}\equiv Q_{\alpha\phantom{\mu}\mu}^{\phantom{\alpha}\mu}\ ,
\end{equation}
and
\begin{equation}\label{16}
\widetilde{Q}^{\alpha}\equiv Q_{\mu}^{\phantom{\mu}\mu\alpha}\ .
\end{equation}
Contracting the Riemann tensor~\eqref{11} as
\begin{equation}\label{16.1}
R=g^{\mu\sigma}\delta_{\lambda}^{\nu}R^{\lambda}_{\phantom{\lambda}\mu\nu\sigma}\ ,
\end{equation}
we obtain, from Eqs.~\eqref{14}--\eqref{16.1},  the curvature scalar $R$ as\footnote{It is worth noticing that some authors define the non-metricity scalar $Q$  as $Q=1/4Q_{\alpha\beta\gamma}Q^{\alpha\beta\gamma}-\dots$ that implies $R=\mathcal{R}+Q+\mathcal{D}_{\alpha}\bigl(Q^{\alpha}-\widetilde{Q}^{\alpha}\bigr)$.}   \label{1000}
 \begin{equation}\label{16.2}
 R=\mathcal{R}-Q+\mathcal{D}_{\alpha}\bigl(Q^{\alpha}-\widetilde{Q}^{\alpha}\bigr)\ ,
 \end{equation}
 where $\mathcal{R}$ and $\mathcal{D}_{\alpha}$ stand for the scalar curvature and the covariant derivative expressed by the LC connection $\hat{\Gamma}^{\alpha}_{\phantom{\alpha}\mu\nu}$ respectively, while $R$ and $Q$ are associated to the STG connection $\Gamma^{\alpha}_{\phantom{\alpha}\mu\nu}$.  In the symmetric teleparallel theories, the curvature vanishes, i.e. $R=0$, and then  Eq.~\eqref{16.2} yields\footnote{If we adopted the previous definition for $Q$, we have $ \mathcal{R}=-Q-B$.}
 \begin{equation}\label{16.3}
 \boxed{
 \mathcal{R}=Q-B
 }\ ,
 \end{equation}
 where $B$ is the boundary term defined as 
 \begin{equation}\label{16.4}
 B=\mathcal{D}_{\alpha}B^{\alpha}\ ,
 \end{equation}
 with vector boundary term $B^{\alpha}$
 \begin{equation}\label{16.5}
 B^{\alpha}=Q^{\alpha}-\widetilde{Q}^{\alpha}\ .
 \end{equation}
 The boundary terms $B$ can be expressed in STG covariant derivative $\nabla$ as 
\begin{equation}\label{47}
B=\nabla_{\alpha}\bigl(Q^{\alpha}-\widetilde{Q}^{\alpha}\bigr)+\frac{1}{2}Q_{\beta}Q^{\beta}-\frac{1}{2}Q_{\beta}\widetilde{Q}^{\beta}\ .
\end{equation}

 \section{The energy and momentum balance equations in  theories of gravity}\label{B}
 Before developing our considerations for obtaining   GW solutions in $f(Q,B)$ gravity, let us define the energy and momentum balance and the deviation equation in generic theories of gravity because these concepts will be useful for the following discussion.
 
The energy and momentum balance equations follow from the covariant divergence of the matter energy-momentum tensor associated to the LC connection which is generally non-zero, i.e., $\mathcal{D}_{\mu}T^{\mu}_{\phantom{\mu}\nu}\neq 0$.  We assume that the matter content of the gravitational system can be described by the energy-momentum tensor of a perfect fluid given by
\begin{equation}\label{50}
T_{\mu\nu}=(\rho +p)u_{\mu}u_{\nu}-p g_{\mu\nu}\ ,
\end{equation}
where $\rho$ and $p$ are the thermodynamic energy density  and isotropic pressure. Here, ${\displaystyle u^{\alpha}=\frac{dx^{\alpha}}{ds}}$ is the four-velocity satisfying the normalization condition $u^{\alpha}u_{\alpha}=1$. Let us now take into account  the covariant divergence of the
energy-momentum tensor~\eqref{50} and project it orthogonally along the four-velocity, thanks to  the projection operator $h^{\lambda}_{\nu}=\delta^{\lambda}_{\nu}-u^{\lambda}u_{\nu}$, which satisfies the following properties: $h^{\lambda}_{\nu}u^{\nu}=0$ and $u_{\alpha}\mathcal{D}_{\nu}h^{\alpha}=0$. By doing so we obtain the  equations of motion of a perfect fluid in any modified theory of gravity, governed by Levi-Civita, where the stress-energy tensor of matter is covariantly non-conserved.  Following~\cite{HKLOR}, it reads
\begin{equation}\label{51}
\boxed{
\frac{d^{2}x^{\lambda}}{ds^{2}}+\hat{\Gamma}^{\lambda}_{\phantom{\lambda}\alpha\beta}\frac{dx^{\alpha}}{ds}\frac{dx^{\beta}}{ds}=\frac{h^{\lambda\nu}\bigl(\mathcal{D}_{\mu}T^{\mu}_{\phantom{\mu}\nu}+\mathcal{D}_{\nu}p\bigr)}{\rho+p}
}\ ,
\end{equation}
or 
\begin{equation}\label{51_1}
\frac{D u^{\lambda}}{ds}=\mathcal{F}^{\lambda}\ ,
\end{equation}
where $\mathcal{F}^{\lambda}$ represents the extra force, and $D/ds$ is the covariant derivative along the parametric curve $x^{\alpha}(s)$ expressed as 
\begin{equation}\label{51_11}
\frac{D}{ds}=u^{\alpha}\mathcal{D}_{\alpha}\ .
\end{equation}
Eq.~\eqref{51} is nothing else but the local version of the {\em momentum balance equation}. Instead, if we contract the covariant divergence of the stress-energy tensor $\mathcal{D}_{\mu}T^{\mu}_{\phantom{\mu}\nu}$ with the four-velocity $u^{\nu}$, we derive the {\em energy balance equation} as 
\begin{equation}\label{52}
\boxed{
\frac{D\rho}{ds}+\left(\rho+p\right)\mathcal{D}_{\mu}u^{\mu}=u^{\nu}\mathcal{D}_{\alpha}T^{\alpha}_{\phantom{\alpha}\nu}=\mathcal{S}
}\ .
\end{equation} 
The right hand side of  Eq.~\eqref{51}, as already mentioned, is the extra force, while the right hand side of Eq.~\eqref{52} can be viewed as a sink or a source term of the energy  $\mathcal{S}$, i.e., the energy creation or annihilation, both due to the non-vanishing of LC covariant divergence of the stress-energy tensor, i.e., $\mathcal{D}_{\mu}T^{\mu}_{\phantom{\mu}\nu}\neq 0$. If the extra force $\mathcal{F}^{\lambda}$  in  Eq.~\eqref{51} is present, then the motion is non-geodesic. This means that test particles do not follow the metric geodesics or, equivalently, the affine geodesics (autoparallel curves) of the Riemannian spacetime with respect to the LC connection. 

\section{The deviation equation  in  theories of gravity}\label{C}
Let us  derive now the deviation equation in arbitrary  theories  of gravity where  matter-energy and momentum tensor are not covariantly  conserved according to  the LC connection, i.e. $\mathcal{D}_{\alpha}T^{\alpha}_{\phantom{\alpha}\beta}\neq 0$.

The presence of  the extra force  induces massive point-like  bodies to move along non-geodesic curves  according to the equation of motion~\eqref{51}.  Now we will obtain the relative acceleration between two nearby worldlines traveled by two free test massive bodies. Then we consider  two nearby free test particles moving on  a  non-metric spacetime manifold.  Their spacetime trajectories are described by  two parametric curves $x^{\mu}(s)$ and $y^{\mu}(s)$, that verify  Eqs.~\eqref{51},  where $s$ is the scalar which parameterizes the trajectory.  The spacelike separation four-vector $\eta^{\mu}$,  defined as
\begin{equation}
y^{\mu}(s)=x^{\mu}(s)+\eta^{\mu}(s)\ ,
\end{equation}
connects points with the same value of $s$ on the two curves.  Let us introduce a perfect fluid consisting of a single particle of rest mass $m$, whose energy-momentum tensor is given by 
\begin{equation}
T^{\mu\nu}=\rho_{0}u^{\mu}u^{\nu}\ ,
\end{equation}
with rest energy density 
\begin{equation}
\rho_{0}(\boldsymbol{x},s)=\frac{m}{\sqrt{-g}}\delta^{3}\left(\boldsymbol{x}-\boldsymbol{x}(s)\right)\ .
\end{equation}
The worldline $x^{\mu}$ satisfies Eq.~\eqref{51} while the wordline $x^{\mu}+\eta^{\mu}$ satisfies 
\begin{equation}\label{51.5}
\frac{d^{2}\left(x^{\mu}+\eta^{\mu}\right)}{ds^{2}}+\hat{\Gamma}^{\mu}_{\phantom{\lambda}\alpha\beta}\left(x+\eta\right)\frac{d\left(x^{\alpha}+\eta^{\alpha}\right)}{ds}\frac{d\left(x^{\beta}+\eta^{\beta}\right)}{ds}=\frac{h^{\mu\nu}\mathcal{D}_{\lambda}T^{\lambda}_{\phantom{\lambda}\nu}}{\rho_{0}}\left(x+\eta\right)\ .
\end{equation}
If $|\eta^{\mu}|$ is much smaller than the typical scale of variation of  gravitational field, taking the difference between Eqs.~\eqref{51.5} and~\eqref{51}, and
expanding up to first order in $\eta$, we get, after some straightforward algebraic calculations, the following equation of motion for the displacement $\eta^{\mu}$ 
\begin{equation}\label{52_2}
\boxed{
\frac{D^{2}\eta^{\mu}}{ds^{2}}=-\mathcal{R}^{\mu}_{\phantom{\mu}\alpha\nu\beta}\eta^{\nu}u^{\alpha}u^{\beta}+\frac{1}{\rho_{0}}\mathcal{D}_{\sigma}\bigl(h^{\nu\mu}\mathcal{D}_{\lambda}T^{\lambda}_{\phantom{\lambda}\nu}\bigr)\eta^{\sigma}
}\ ,
\end{equation}
according to the definition of the covariant derivative of a vector along the curve $x^{\mu}(s)$ Eq.~\eqref{51_11}. In Eq.~\eqref{52_2} the Riemann tensor $\mathcal{R}^{\mu}_{\phantom{\mu}\alpha\nu\beta}$ is expressed in terms of the LC connection $\hat{\Gamma}^{\mu}_{\phantom{\lambda}\alpha\beta}$. in metric spaces, the second term in the r.h.s. disappears.

\section{$f(Q,B)$ non-metric gravity}\label{D}
We consider now a non-linear extension of STEGR,  where an arbitrary analytic function $f$ of  the non-metricity scalar $Q$ and the  boundary term $B$,  previously defined in Eq.~\eqref{16.4} is taken into account.  Clearly this action is more general than $f(Q)$ non-metric theories \cite{LaHe}. Hence, the action of  $f(Q,B)$ theories of gravity can be expressed as~\cite{Carmen,SFSGS} 
\begin{equation}\label{13_1}
S=\int_{\Omega}d^{4}x\,\Bigl[\frac{1}{2\kappa^{2}}\sqrt{-g}f\left(Q,B\right)+\lambda_{\alpha}^{\phantom{\alpha}\beta\mu\nu}R^{\alpha}_{\phantom{\alpha}\beta\mu\nu}+\lambda_{\alpha}^{\phantom{\alpha}\mu\nu}T^{\alpha}_{\phantom{\alpha}\mu\nu}+\sqrt{-g}\mathcal{L}_{m}\bigl(g\bigr)\Bigr]\ ,
\end{equation}
where $\kappa^{2}=8\pi G/c^{4}$ and $\lambda_{\alpha}^{\phantom{\alpha}\beta\mu\nu}=\lambda_{\alpha}^{\phantom{\alpha}\beta[\mu\nu]}$, $\lambda_{\alpha}^{\phantom{\alpha}\mu\nu}=\lambda_{\alpha}^{\phantom{\alpha}[\mu\nu]}$ are the Lagrange multipliers needed to impose the two conditions of curvature-free and torsion-free~\eqref{11} and \eqref{12}, i.e.  further 96 plus 24 independent scalar fields.

\subsection{The variation of the action. Field and connection equations}\label{DA}
First, we need to introduce a new very useful geometric object, the non-metricity conjugate tensor defined as 
\begin{equation}\label{17}
P^{\alpha}_{\phantom{\alpha}\mu\nu}=\frac{1}{2}\frac{\partial Q}{\partial Q_{\alpha}^{\phantom{\alpha}\mu\nu}}\ ,
\end{equation}
which is  symmetric in the last two indices
\begin{equation}
P^{\alpha}_{\phantom{\alpha}\mu\nu}=P^{\alpha}_{\phantom{\alpha}(\mu\nu)}\ ,
\end{equation}
and allows us to write the non-metricity scalar Q as
\begin{equation}\label{18}
Q=Q_{\alpha\mu\nu}P^{\alpha\mu\nu}\ .
\end{equation}
Therefore this superpotential $P^{\alpha}_{\phantom{\alpha}\mu\nu}$ can be written explicitly as 
\begin{eqnarray}\label{24}
P^{\alpha}_{\phantom{\alpha}\mu\nu}&=&\frac{1}{4}\Bigl[-Q^{\alpha}_{\phantom{\alpha}\mu\nu}+2Q_{(\mu\phantom{\alpha}\nu)}^{\phantom{(\mu}\alpha}+Q^{\alpha}g_{\mu\nu}-\widetilde{Q}^{\alpha}g_{\mu\nu}-\delta^{\alpha}_{(\mu}Q_{\nu)}\Bigr]\nonumber\\
&=&-\frac{1}{2}L^{\alpha}_{\phantom{\alpha}\mu\nu}+\frac{1}{4}\bigl(Q^{\alpha}-\widetilde{Q}^{\alpha}\big)g_{\mu\nu}-\frac{1}{4}\delta^{\alpha}_{(\mu}Q_{\nu)}\ .
\end{eqnarray}
Then, in the Palatini formalism, performing the variation with respect to the metric tensor $g_{\mu\nu}$ of the action~\eqref{13_1},  we obtain 
\begin{equation}\label{80_5}
\delta_{g}S=\int_{\Omega}d^{4}x\,\Biggl\{\frac{1}{2\kappa^{2}}\biggl[\delta_{g}\Bigl(\sqrt{-g}\Bigr)f(Q,B)+\sqrt{-g}\Bigl(f_{Q}\delta_{g}Q+f_{B}\delta_{g}B\Bigr)\biggr]+\delta_{g}\Bigl[\sqrt{-g}\mathcal{L}_{m}\bigl(g\bigr)\Bigr]\Biggr\}\ .
\end{equation}
Assuming that the fields vanish on the domain boundary  to cancel the surface integrals and by the variations \eqref{8.9}, \eqref{8.10} and \eqref{8.11} in appendix \ref{F_5}, the following expression 
\begin{equation}\label{80_8}
\int_{\Omega}d^{4}x\,\sqrt{-g}f_{B}B^{\alpha}\delta_{g}\hat{\Gamma}^{\sigma}_{\phantom{\sigma}\alpha\sigma}=\int_{\Omega}d^{4}x\,\Bigl[\frac{1}{2}\sqrt{-g}g_{\mu\nu}f_{B}B+\frac{1}{2}\sqrt{-g}\mathcal{D}_{\alpha}f_{B}\bigl(Q^{\alpha}-\widetilde{Q}^{\alpha}\bigr)g_{\mu\nu}\Bigr]\delta g^{\mu\nu}\ ,
\end{equation}
and the independence and symmetry of the $\delta g^{\mu\nu}$, we get  the metric field equation of $f(Q,B)$ non-metric gravity, i.e.,
\begin{equation}\label{82}
\boxed{
\begin{aligned}
\frac{2}{\sqrt{-g}}\nabla_{\alpha}\left(\sqrt{-g}f_{Q}P^{\alpha}_{\phantom{\alpha}\mu\nu}\right)&-\frac{1}{2}g_{\mu\nu}f+f_{Q}\Bigl(P_{\mu\alpha\beta}Q_{\nu}^{\phantom{\nu}\alpha\beta}-2Q^{\alpha\beta}_{\phantom{\alpha\beta}\mu}P_{\alpha\beta\nu}\Bigr)\\
&+\frac{1}{2}g_{\mu\nu}f_{B}B+2\mathcal{D}_{\alpha}f_{B}P^{\alpha}_{\phantom{\alpha}\mu\nu}+\mathcal{D}_{\mu}\mathcal{D}_{\nu}f_{B}-g_{\mu\nu}\mathcal{D}^{\alpha}\mathcal{D}_{\alpha}f_{B}=\kappa^{2}T_{\mu\nu}
\end{aligned}
}\ ,
\end{equation}
where $f_{Q}=\partial f/\partial Q$, $f_{B}=\partial f/\partial B$, and the matter energy-momentum tensor is
\begin{equation}\label{23}
T_{\mu\nu}=-\frac{2}{\sqrt{-g}}\frac{\delta\bigl(\sqrt{-g}\mathcal{L}_{m}\bigr)}{\delta g^{\mu\nu}}\ .
\end{equation}
If we raise the first covariant index of Eq.~\eqref{82}, we obtain the following mixed $(1,1)$ form 
\begin{equation}\label{83}
\boxed{
\begin{aligned}
\frac{2}{\sqrt{-g}}\nabla_{\alpha}&\left(\sqrt{-g}f_{Q}P^{\alpha\mu}_{\phantom{\alpha\mu}\nu}\right)-\frac{1}{2}\delta^{\mu}_{\nu}f+f_{Q}P^{\mu}_{\phantom{\mu}\alpha\beta}Q_{\nu}^{\phantom{\nu}\alpha\beta}\\
&+\frac{1}{2}\delta^{\mu}_{\nu}f_{B}B+2\mathcal{D}_{\alpha}f_{B}P^{\alpha\mu}_{\phantom{\alpha\mu}\nu}+\mathcal{D}^{\mu}\mathcal{D}_{\nu}f_{B}-\delta^{\mu}_{\nu}\mathcal{D}^{\alpha}\mathcal{D}_{\alpha}f_{B}=\kappa^{2}T^{\mu}_{\phantom{\mu}\nu}
\end{aligned}
}\ .
\end{equation}
On the other hand,  the variation of  action~\eqref{13_1} with respect to the affine connection $\Gamma$ leads to  
\begin{equation}
\delta_{\Gamma}S=\int_{\Omega}d^{4}x\,\Biggl\{\frac{\sqrt{-g}}{2\kappa^{2}}\Bigl(f_{Q}\delta_{\Gamma}Q+f_{B}\delta_{\Gamma}B\Bigr)+\lambda_{\alpha}^{\phantom{\alpha}\beta\mu\nu}\delta_{\Gamma}R^{\alpha}_{\phantom{\alpha}\beta\mu\nu}+\lambda_{\alpha}^{\phantom{\alpha}\mu\nu}\delta_{\Gamma}T^{\alpha}_{\phantom{\alpha}\mu\nu}\Biggr\}\ ,
\end{equation}
because 
\begin{equation}
\delta_{\Gamma}\sqrt{-g}=\delta_{\Gamma}\mathcal{L}_{m}(g)=\delta_{\Gamma}\lambda_{\alpha}^{\phantom{\alpha}\beta\mu\nu}=\delta_{\Gamma}\lambda_{\alpha}^{\phantom{\alpha}\mu\nu}=0\ .
\end{equation}
Therefore, the stationary condition $\delta_{\Gamma}S=0$ according to the variations in appendix \ref{F_5},  Eqs.~\eqref{8.60}--\eqref{8.62} and \eqref{8.66}, from the arbitrariness and symmetry in two latest indices of the $\delta\Gamma^{\alpha}_{\phantom{\alpha}\mu\nu}$, lead to 
\begin{equation}\label{85}
2\kappa^{2}\Bigl(\nabla_{\beta}\lambda_{\alpha}^{\phantom{\alpha}\nu\mu\beta}+\lambda_{\alpha}^{\phantom{\alpha}\mu\nu}\Bigr)=2\sqrt{-g}f_{Q}P^{(\mu\nu)}_{\phantom{\mu\nu}\alpha}+\frac{\sqrt{-g}}{2}\Bigl(g^{\mu[\nu}\delta_{\alpha}^{\rho]}+g^{\nu[\mu}\delta_{\alpha}^{\rho]}\Bigr)\mathcal{D}_{\rho}f_{B}\ .
\end{equation}\label{86}
Thanks to the commutativity of the two STG covariant derivatives, $\nabla_{\mu}$ and $\nabla_{\nu}$,  the curvature and the torsion of STG connection  vanishing, taking into account the following equalities 
\begin{equation}\label{86_5}
2\sqrt{-g}P^{(\lambda\sigma)}_{\phantom{\lambda\sigma}\nu}=-\nabla_{\omega}\Bigl(\sqrt{-g}\,\delta^{(\lambda}_{\rho}g^{\sigma)[\rho}\delta^{\omega]}_{\nu}\Bigr)\ ,
\end{equation}
and 
\begin{equation}
\nabla_{\mu}\nabla_{\nu}\Bigl[\sqrt{-g}\,\delta^{\lambda}_{\rho}g^{\mu[\nu}\delta_{\alpha}^{\rho]}\nabla_{\lambda}f_{B}\Bigr]=-\nabla_{\mu}\nabla_{\nu}\Bigl[\nabla_{\lambda}\Bigl(\sqrt{-g}\,\delta^{\lambda}_{\rho}g^{\mu[\nu}\delta_{\alpha}^{\rho]}\Bigr)f_{B}\Bigr]\ ,
\end{equation}
we finally find the {\em STG connection equation of motion in $f(Q,B)$ non-metric gravity}
\begin{equation}\label{87}
\boxed{
\nabla_{\mu}\nabla_{\nu}\Bigl[\sqrt{-g}\bigl(f_{Q}+f_{B}\bigr)P^{\mu\nu}_{\phantom{\mu\nu}\alpha}\Bigr]=0
}\ .
\end{equation}
For a  similar treatment in $f(Q)$, see references~\cite{JHKP,DZ,DKC, CDA, BVC, CFSM, VCCE, ABS, Mehdi}
\subsection{The Levi-Civita covariant divergence of  matter energy-momentum tensor}\label{DB}
Now we explicitly calculate the LC covariant divergence of the matter energy-momentum tensor $T^{\mu}_{\phantom{\mu}\nu}$ in $f(Q,B)$ theories and verify that it vanish, that is $\mathcal{D}_{\mu}T^{\mu}_{\phantom{\mu}\nu}=0$ on-shell, after performing the LC covariant divergence $\mathcal{D}_{\mu}H^{\mu}_{\phantom{\mu}\nu}$.  Like in $f(Q)$ non-metric gravity, both  energy and momentum densities are LC covariantly conserved in  symmetric teleparallel non-metric theories with boundary term.  Hence, to prove it, we first observe that the geometric object $\mathcal{E}^{\mu}_{\phantom{\mu}\nu}=\sqrt{-g}T^{\mu}_{\phantom{\mu}\nu}$ is a mixed tensor density of weight 1 and therefore its STG covariant derivative is given by~\eqref{8.21}
\begin{equation}
\nabla_{\alpha}\mathcal{E}^{\alpha}_{\phantom{\alpha}\nu}=\mathcal{D}_{\alpha}\mathcal{E}^{\alpha}_{\phantom{\alpha}\nu}-L^{\lambda}_{\phantom{\lambda}\alpha\nu}\mathcal{E}^{\alpha}_{\phantom{\alpha}\lambda}\ .
\end{equation}
Considering the non-metricity of connection, it gives
\begin{equation}\label{37.2}
\nabla_{\mu}P^{\mu\lambda}_{\phantom{\mu\lambda}\alpha}=\nabla_{\mu}P^{\mu\lambda\beta}+Q_{\mu\alpha\beta}P^{\mu\lambda\beta}\ .
\end{equation}
From the symmetry in the last two indices of $P_{\alpha\beta\gamma}$ and the definition of  disfomation tensor $L_{\alpha\beta\gamma}$, it  follows that
\begin{equation}\label{37.4}
\left(Q_{\nu\alpha\beta}+2L_{\beta\alpha\nu}\right)P^{\mu\alpha\beta}=\left(Q_{\nu\alpha\beta}+2L_{\beta\alpha\nu}\right)\nabla_{\mu}P^{\mu\alpha\beta}=0\ .
\end{equation}
Hence, from the Bianchi identities~\eqref{8.8} for the non-metricity tensor $Q_{\alpha\mu\nu}$ and from Eqs.~\eqref{8.17} and \eqref{8.18} in Appendix, from the previous identities~\eqref{37.4}, together with the symmetry in the last indices of  $L_{\alpha\beta\gamma}$, it yields 
\begin{equation}\label{37.5}
f_{Q}\left(\nabla_{\mu}Q_{\nu\alpha\beta}+L^{\rho}_{\phantom{\rho}\mu\nu}Q_{\rho\alpha\beta}+2L^{\rho}_{\phantom{\rho}\alpha\nu}Q_{\mu\beta\rho}\right)P^{\mu\alpha\beta}=f_{Q}\left(\mathcal{D}_{\nu}Q_{\mu\alpha\beta}\right)P^{\mu\alpha\beta}\ .
\end{equation}
If we use  definitions of the non-metricity conjugate tensor or the superpotential $P^{\alpha}_{\phantom{\alpha}\mu\nu}$~\eqref{24} and the non-metricity scalar $Q$~\eqref{14}, we obtain 
\begin{equation}\label{37.6}
Q_{\mu\alpha\beta}\mathcal{D}_{\nu}P^{\mu\alpha\beta}=\frac{1}{2}\mathcal{D}_{\nu}Q\ .
\end{equation}
Finally from Eqs.~\eqref{37.5} and \eqref{37.6}, we derive 
\begin{equation}\label{37.7}
f_{Q}\left(\mathcal{D}_{\nu}Q_{\mu\alpha\beta}\right)P^{\mu\alpha\beta}=\frac{1}{2}f_{Q}\nabla_{\nu}Q\ .
\end{equation}
Let us now perform the covariant divergence on both sides of the field Eqs.~\eqref{83} associated to the LC connection, and according to Eqs.~\eqref{37.2}--\eqref{37.7}, we get
\begin{equation}\label{500}
\kappa^{2}\mathcal{D}_{\mu}T^{\mu}_{\phantom{\mu}\nu}=\frac{2}{\sqrt{-g}}\nabla_{\mu}\nabla_{\alpha}\Bigl(\sqrt{-g}f_{Q}P^{\alpha\mu}_{\phantom{\alpha\mu}\nu}\bigr)+\frac{1}{2}\nabla_{\nu}f_{B}B+2\mathcal{D}_{\mu}\Bigl(\nabla_{\alpha}f_{B}P^{\alpha\mu}_{\phantom{\alpha\mu}\nu}\Bigr)+\bigl[\mathcal{D}^{\alpha},\mathcal{D}_{\nu}\bigr]\nabla_{\alpha}f_{B}\ .
\end{equation}
Taking into account that curvature and torsion of non-metric compatible STG connection vanish in  symmetric teleparallel theories, we obtain the following expression for the Riemann tensor $\mathcal{R}^{\alpha}_{\phantom{\alpha}\beta\mu\nu}$ (in the LC connection) in terms of the disformation tensor $L^{\alpha}_{\phantom{\alpha}\mu\beta}$, namely 
\begin{equation}\label{510}
\mathcal{R}^{\alpha}_{\phantom{\alpha}\beta\mu\nu}=\nabla_{\nu}L^{\alpha}_{\phantom{\alpha}\mu\beta}-\nabla_{\mu}L^{\alpha}_{\phantom{\alpha}\nu\beta}+L^{\alpha}_{\phantom{\alpha}\mu\lambda}L^{\lambda}_{\phantom{\lambda}\nu\beta}-L^{\alpha}_{\phantom{\alpha}\nu\lambda}L^{\lambda}_{\phantom{\lambda}\mu\beta}\ .
\end{equation}
From which, contracting the first and the third index and raising the second covariant index, we get the Ricci tensor in LC connection, that is
\begin{multline}\label{515}
\mathcal{R}^{\beta}_{\phantom{\beta}\nu}=L^{\alpha}_{\phantom{\alpha}\nu\lambda}L^{\lambda\beta}_{\phantom{\lambda\beta}\alpha}-L^{\alpha}_{\phantom{\alpha}\alpha\lambda}L^{\lambda\beta}_{\phantom{\lambda\beta}\nu}+\mathcal{D}_{\nu}L^{\alpha\beta}_{\phantom{\lambda\beta}\alpha}-\mathcal{D}_{\alpha}L^{\alpha\beta}_{\phantom{\alpha\beta}\nu}\\
=\frac{2}{\sqrt{-g}}\nabla_{\lambda}\left(\sqrt{-g}P^{\lambda\beta}_{\phantom{\lambda\beta}\nu}\right)+P^{\beta}_{\phantom{\beta}\omega\lambda}Q_{\nu}^{\phantom{\nu}\omega\lambda}-\frac{1}{2}\mathcal{D}_{\omega}\left(Q^{\omega}-\widetilde{Q}^{\omega}\right)\delta^{\beta}_{\nu}\ .
\end{multline}
According to  Eq.~\eqref{515}, the commutator of LC covariant derivatives yields
\begin{equation}\label{516}
\bigl[\mathcal{D}^{\alpha},\mathcal{D}_{\nu}\bigr]\nabla_{\alpha}f_{B}=\mathcal{R}^{\beta}_{\phantom{\beta}\nu}\nabla_{\beta}f_{B}\ ,
\end{equation}
and together with Eqs.~\eqref{87} and \eqref{515}, it allow us to express the four-divergence~\eqref{500} as 
\begin{equation}\label{520}
\begin{aligned}
\kappa^{2}\mathcal{D}_{\mu}T^{\mu}_{\phantom{\mu}\nu}=-\frac{2}{\sqrt{-g}}\nabla_{\mu}\nabla_{\alpha}\Bigl(\sqrt{-g}P^{\alpha\mu}_{\phantom{\alpha\mu}\nu}\Bigr)f_{B}\ .
\end{aligned}
\end{equation}\label{521}
Finally, from  Eq.~\eqref{86_5} and because the covariant STG derivatives commute, it is 
\begin{equation}
2\nabla_{\mu}\nabla_{\alpha}\Bigl(\sqrt{-g}P^{\alpha\mu}_{\phantom{\alpha\mu}\nu}\Bigr)=\nabla_{\mu}\nabla_{\alpha}\nabla_{\omega}\Bigl(\sqrt{-g}\,\delta^{(\alpha}_{\rho}g^{\mu)[\omega}\delta_{\nu}^{\rho]}\Bigr)=0\ ,
\end{equation}
that leads to the LC covariant conservation of the matter energy-momentum tensor in $f(Q,B)$ gravity on-shell, namely 
\begin{equation}\label{522}
\boxed{
\mathcal{D}_{\mu}T^{\mu}_{\phantom{\mu}\nu}=0
}\ .
\end{equation}
\subsection{The  equation of motion of a massive particle and the deviation equation}\label{DC}
We want to derive now  the  equation of motion of a test mass and the deviation equation in $f(Q,B)$ non-metric gravity. From the Eq.\eqref{522},   free structure-less particles follow the geodesic equations 
\begin{equation}\label{600}
\frac{d^{2}x^{\lambda}}{d\tau^{2}}+\hat{\Gamma}^{\lambda}_{\phantom{\lambda}\alpha\beta}\frac{dx^{\alpha}}{d\tau}\frac{dx^{\beta}}{d\tau}=0\ ,
\end{equation}
according to Eq.~\eqref{51} where we set the divergence $\mathcal{D}_{\mu}T^{\mu}_{\phantom{\mu}\nu}$ equal to zero and  $\tau$ is the proper time.  While the deviation equation for a theory  where  energy and  momentum densities are LC covariantly conserved,  Eq.~\eqref{52_2} becomes 
\begin{equation}\label{601}
\frac{D^{2}\eta^{\mu}}{d\tau^{2}}=-\mathcal{R}^{\mu}_{\phantom{\mu}\alpha\nu\beta}\eta^{\nu}u^{\alpha}u^{\beta}\ ,
\end{equation}
that is, the well-known geodesic deviation equation of  GR. In terms of the symmetric teleparallel connection, Eqs.~\eqref{600} and~\eqref{601} become, respectively
\begin{equation}\label{602}
\boxed{
\frac{D^{\prime}u^{\nu}}{d\tau}=L^{\nu}_{\phantom{\nu}\alpha\beta}u^{\alpha}u^{\beta}}\ ,
\end{equation}
or equivalently 
\begin{equation}\label{603}
\frac{d^{2}x^{\lambda}}{d\tau^{2}}+\Gamma^{\lambda}_{\phantom{\lambda}\alpha\beta}\frac{dx^{\alpha}}{d\tau}\frac{dx^{\beta}}{d\tau}=L^{\lambda}_{\phantom{\nu}\alpha\beta}u^{\alpha}u^{\beta}\ ,
\end{equation}
and 
\begin{equation}\label{604}
\boxed{
\frac{{D^{\prime}}^{2}\eta^{\nu}}{d\tau^2}-2L^{\nu}_{\phantom{\nu}\alpha\beta}\frac{D^{\prime}\eta^{\alpha}}{d\tau}u^{\beta}=\nabla_{\sigma}L^{\nu}_{\phantom{\nu}\alpha\beta}\eta^{\sigma}u^{\alpha}u^{\beta}}\ ,
\end{equation}
with LC and STG covariant derivative along the curve $x^{\lambda}=x^{\lambda}(\tau)$ with tangent vector $u^{\alpha}=dx^{\alpha}/d\tau$, defined as  
\begin{equation}\label{605}
\frac{D}{d\tau}=u^{\lambda}\mathcal{D}_{\lambda}\ ,
\end{equation}
and 
\begin{equation}
\frac{D^{\prime}}{d\tau}=u^{\lambda}\nabla_{\lambda}\ ,
\end{equation}
respectively. 

\section{Linearized field and connection equations in vacuum}\label{DD}
We have now all the ingredients to linearize the field and connection equations and to derive GW solutions. We will perform these calculations for coincident gauge and for any gauge. We  show the complete independence of the theory from the gauge choice.

\subsection{ The case of  coincident gauge}
It is well know that,  in STG, it is always possible to trivialize the connection, that is, we can choose a coordinate system such that 
\begin{equation}\label{19}
\Gamma^{\alpha}_{\phantom{\alpha}\mu\nu}=0\ ,
\end{equation}
i.e., the so-called {\it coincident gauge}.  The flatness of  connection~\eqref{11} makes it integrable and therefore it can be written as
\begin{equation}
\Gamma^{\alpha}_{\phantom{\alpha}\mu\nu}=\left(\Lambda^{-1}\right)^{\alpha}_{\phantom{\mu}\gamma}\partial_{\mu}\Lambda^{\gamma}_{\phantom{\gamma}\nu}\ ,
\end{equation}
with the matrices $\Lambda^{\gamma}_{\phantom{\gamma}\nu}\in \text{GL}(4,\mathbb{R})$,  i.e. belonging to the general linear group. The absence of torsion ~\eqref{12} gives 
\begin{equation}
\partial_{[\beta}\Lambda^{\gamma}_{\phantom{\gamma}\nu]}=0\ .
\end{equation}
This implies that the transformation can be parameterized with a vector $\xi^{\mu}$ as
\begin{equation}
\Lambda^{\mu}_{\phantom{\mu}\nu}=\partial_{\nu}\xi^{\mu}\ .
\end{equation}
Finally the connection can be given as
\begin{equation}
\Gamma^{\alpha}_{\phantom{\alpha}\mu\nu}=\frac{\partial x^{\alpha}}{\partial\xi^{\beta}}\partial_{\mu}\partial_{\nu}\xi^{\beta}\ .
\end{equation}
So it is always possible to choose a coordinate system such that $x^{\mu}=\xi^{\mu}$, and then the connection vanishes.  Physically, this means that the origin of the tangent space coincides with the origin of the spacetime. Hence, in concident gauge, we have 
\begin{equation}\label{20}
\hat{\Gamma}^{\alpha}_{\phantom{\alpha}\mu\nu}=-L^{\alpha}_{\phantom{\alpha}\mu\nu}\ ,
\end{equation}
and  the non-metricity tensor becomes 
\begin{equation}\label{21}
Q_{\alpha\mu\nu}=\partial_{\alpha}g_{\mu\nu}\ ,
\end{equation}
that is,  the covariant derivative $\nabla$ associated to the connection $\Gamma$ becomes the partial derivative $\partial$.
It is worth noticing that  $\nabla_{\alpha}g_{\mu\nu}\neq 0$,  while only $\mathcal{D}_{\alpha}g_{\mu\nu}=0$, where $\mathcal{D}_{\alpha}$, as already mentioned, represents the covariant derivative associated to the LC connection $\hat{\Gamma}^{\alpha}_{\phantom{\alpha}\mu\nu}$.
So, we adopt the coincident gauge and weakly perturb  the metric tensor $g_{\mu\nu}$ around a Minkowsky background as 
\begin{equation}\label{25_0}
g_{\mu\nu}=\eta_{\mu\nu}+h_{\mu\nu}\ .
\end{equation}
Thus, in coincident gauge,  field Eqs.~\eqref{83} to the first order in the perturbation $h_{\mu\nu}$ in vacuum reads as 
\begin{multline}\label{60}
-\frac{1}{2}f_{Q}(0,0)\Bigl[\Box h_{\mu\nu}-\bigl(\partial_{\alpha}\partial_{\mu}h^{\alpha}_{\phantom{\alpha}\nu}+\partial_{\alpha}\partial_{\nu}h^{\alpha}_{\phantom{\alpha}\mu}\bigr)-\eta_{\mu\nu}\bigl(\Box h-\partial_{\alpha}\partial_{\beta}h^{\alpha\beta}\bigr)+\partial_{\mu}\partial_{\nu}h\Bigl]\\
+f_{BB}(0,0)\Bigl[\eta_{\mu\nu}\bigl(\Box\partial_{\alpha}\partial_{\beta}h^{\alpha\beta}-\Box^{2}h\bigr)-\partial_{\mu}\partial_{\nu}\partial_{\alpha}\partial_{\beta}h^{\alpha\beta}+\partial_{\mu}\partial_{\nu}\Box h\Bigr]=0\ .
\end{multline}
Here we expanded the function $f(Q,B)$ around the point $(0,0)$, that is, around the values of $Q$ and $B$ on the unperturbed background with the metric $\eta_{\mu\nu}$, as 
\begin{multline}\label{61}
f(Q,B)=f(0,0)+f_{Q}(0,0)Q+f_{B}(0,0)B+\frac{1}{2}f_{QQ}(0,0)Q^{2}\\
+\frac{1}{2}f_{BB}(0,0)B^2+f_{QB}(0,0)QB+\mathcal{O}(Q^3, B^3,\dots)\ ,
\end{multline}
from which one has
\begin{equation}
f^{(0)}=f(0,0)=0\ ,\quad f^{(1)}=f_{B}(0,0)B^{(1)}\ .
\end{equation}
From the following expansions 
\begin{equation}
f_{B}(Q,B)=f_{B}(0,0)+f_{BB}(0,0)B+O(2)\ ,
\end{equation}
\begin{equation}
f_{BB}(Q,B)=f_{BB}(0,0)+O(1)\ ,
\end{equation}
\begin{equation}
f_{Q}(Q,B)=f_{Q}(0,0)+f_{QB}(0,0)B+O(2)\ ,
\end{equation}
we get 
\begin{equation}
f_{Q}^{(0)}=f_{Q}(0,0)\ ,\quad f_{B}^{(0)}=f_{B}(0,0)\ ,\quad f_{BB}^{(0)}=f_{BB}(0,0)\ ,\quad{\mathcal{D}_{\alpha}f_{B}}^{(0)}=0\ .
\end{equation}
As a consequence,  the trace of  field Eqs.~\eqref{60} becomes  
\begin{equation}\label{63}
\boxed{
3f_{BB}(0,0)\bigl(\Box\partial_{\alpha}\partial_{\beta}h^{\alpha\beta}-\Box^{2}h\bigr)+f_{Q}(0,0)\bigl(\Box h-\partial_{\alpha}\partial_{\beta}h^{\alpha\beta}\bigr)=0
}\ .
\end{equation}
In coincidence gauge,  the boundary term $B$ is written as  
\begin{equation}
B=\mathcal{D}_{\sigma}B^{\sigma}=g^{\sigma\rho}\mathcal{D}_{\sigma}\hat{\Gamma}^{\nu}_{\phantom{\nu}\nu\rho}-g^{\mu\nu}\mathcal{D}_{\sigma}\hat{\Gamma}^{\sigma}_{\phantom{\sigma}\mu\nu}\ ,
\end{equation}
with $B^{\sigma}$
\begin{equation}
B^{\sigma}=Q^{\sigma}-\widetilde{Q}^{\sigma}=g^{\sigma\rho}\bigl(\partial_{\rho}g_{\mu\nu}\bigr)g^{\mu\nu}-g^{\sigma\rho}\bigl(\partial_{\mu}g_{\rho\nu}\bigr)g^{\mu\nu}\ .
\end{equation}
Therefore, in coincident gauge, the boundary term $B$ to the first order in $h_{\mu\nu}$ is 
\begin{equation}\label{62}
\tilde{B}^{(1)}=\Box h- \partial^{\alpha}\partial^{\beta}h_{\alpha\beta}\ ,
\end{equation}
 and from Eq.~\eqref{63}, it is possible derive a Klein-Gordon (KG)  equation  for  $\tilde{B}^{(1)}$ as
\begin{equation}\label{62_KG}
\boxed{
\bigl[\Box+m_{B}^{2}\bigr]\tilde{B}^{(1)}=0
}\ ,
\end{equation}
where 
\begin{equation}\label{63_0}
m_{B}^{2}=-\frac{f_{Q}(0,0)}{3f_{BB}(0,0)}\ ,
\end{equation}
that is, $m_{B}^2$ is the square of the effective mass associated to the scalar mode of  linear perturbation due to the boundary term $\tilde{B}^{(1)}$.
Thanks to the trace Eq.~\eqref{63},  field Eqs.~\eqref{60} reduce to
\begin{equation}\label{64}
\boxed{
\begin{aligned}
&f_{BB}(0,0)\Bigl[\partial_{\mu}\partial_{\nu}\Box h-\partial_{\mu}\partial_{\nu}\partial_{\alpha}\partial_{\beta}h^{\alpha\beta}\Bigr]\\
-\frac{1}{2}f_{Q}(0,0)\Bigl[\Box h_{\mu\nu}&-\bigl(\partial_{\alpha}\partial_{\mu}h^{\alpha}_{\phantom{\alpha}\nu}+\partial_{\alpha}\partial_{\nu}h^{\alpha}_{\phantom{\alpha}\mu}\bigr)-\frac{1}{3}\eta_{\mu\nu}\bigl(\Box h-\partial_{\alpha}\partial_{\beta}h^{\alpha\beta}\bigr)+\partial_{\mu}\partial_{\nu}h\Bigl]=0
\end{aligned}
}\ .
\end{equation}
The linearized connection equation, from Eq.~\eqref{87}, becomes
\begin{equation}\label{65_4}
\partial_{\mu}\partial_{\nu}{\widetilde{P^{\mu\nu}_{\phantom{\mu\nu}\alpha}}}^{(1)}=0\ ,
\end{equation}
where ${\widetilde{P^{\mu\nu}_{\phantom{\mu\nu}\alpha}}}^{(1)}$ is the first-order non-metricity conjugate tensor in the metric perturbation $h$ selected in coincidence gauge. Since Eqs.~\eqref{65_4} add no further constraints to  Eqs.~\eqref{64}, we will not consider it.
\subsection{ The case of  any  gauge}
We rewrite now the linearized metric field and connection Eqs.~\eqref{64} and \eqref{65_4} without fixing the gauge. To linearize the non-metric gravity describes by  field Eqs.~\eqref{83}, we perturb around the Minkowski spacetime both metric tensor and connection to first order, considered independent \cite{CC}. It is  
\begin{equation}\label{40_5}
g_{\mu\nu}=\eta_{\mu\nu}+h_{\mu\nu}\ ,\quad\text{and}\quad\Gamma^{\alpha}_{\phantom{\alpha}\mu\nu}={\Gamma^{\alpha}}^{(0)}_{\mu\nu}+{\Gamma^{\alpha}}^{(1)}_{\mu\nu}\ ,
\end{equation}
where $\vert h_{\mu\nu}\vert \ll 1$ and $\vert{\Gamma^{\alpha}}^{(1)}_{\mu\nu}\vert\ll 1$.  Because the non-metric connection disappears to zero-order when gravity is absent, we have
\begin{equation}\label{41_5}
{\Gamma^{\alpha}}^{(0)}_{\mu\nu}=0\ ,
\end{equation}
which reproduces flat spacetime. In any gauge, to the first order in metric and connection perturbations, we get the following linear corrections 
\begin{equation}\label{42_5}
Q_{\alpha\mu\nu}^{(1)}=\partial_{\alpha}h_{\mu\nu}-2\,\Gamma^{(1)}_{(\mu|\alpha|\nu)}\ ,
\end{equation}
\begin{equation}\label{43_5}
{Q^{\alpha}}^{(1)}=\partial^{\alpha}h-2\,{\Gamma^{\lambda\phantom{\lambda}\alpha}_{\phantom{\lambda}\lambda}}^{(1)}\ ,
\end{equation}
\begin{equation}\label{44_5}
{\widetilde{Q}}^{\alpha(1)}=\partial_{\beta}h^{\alpha\beta}-2\,{\Gamma^{(\beta\phantom{\beta}\alpha)}_{\phantom{\beta}\beta}}^{(1)}\ ,
\end{equation}
\begin{equation}\label{45_5}
{B^{\alpha}}^{(1)}={Q^{\alpha}}^{(1)}-{\widetilde{Q}}^{\alpha(1)}=\partial^{\alpha}h-\partial_{\beta}h^{\alpha\beta}+{\Gamma^{\alpha\phantom{\beta}\beta}_{\phantom{\alpha}\beta}}^{(1)}-{\Gamma^{\beta\phantom{\beta}\alpha}_{\phantom{\beta}\beta}}^{(1)}\ .
\end{equation}
The linearized disformation tensor $L^{\alpha}_{\phantom{\alpha}\mu\nu}$ becomes
\begin{equation}\label{46_5}
{L^{\alpha}_{\phantom{\alpha}\mu\nu}}^{(1)}=\frac{1}{2}\partial^{\alpha}h_{\mu\nu}-\partial_{(\mu}h^{\alpha}_{\phantom{\alpha}\nu)}+{\Gamma^{\alpha}_{\phantom{\alpha}\mu\nu}}^{(1)}\ ,
\end{equation}
while the linearized non-metricity conjugate tensor $P^{\alpha}_{\phantom{\alpha}\mu\nu}$ yields 
\begin{multline}\label{47_5}
{P^{\alpha}_{\phantom{\alpha}\mu\nu}}^{(1)}=-\frac{1}{4}\partial^{\alpha}h_{\mu\nu}+\frac{1}{2}\partial_{(\mu}h^{\alpha}_{\phantom{\alpha}\nu)}+\frac{1}{4}\bigl(\partial^{\alpha}h-\partial_{\beta}h^{\alpha\beta}\bigr)\eta_{\mu\nu}-\frac{1}{4}\delta^{\alpha}_{(\mu}\partial_{\nu)}h\\
+\frac{1}{4}\bigl({\Gamma^{\alpha\phantom{\beta}\beta}_{\phantom{\alpha}\beta}}^{(1)}-{\Gamma^{\beta\phantom{\beta}\alpha}_{\phantom{\beta}\beta}}^{(1)}\bigr)\eta_{\mu\nu}-\frac{1}{2}{\Gamma^{\alpha}_{\phantom{\alpha}\mu\nu}}^{(1)}+\frac{1}{2}\delta^{\alpha}_{(\mu}{\Gamma^{\lambda}_{\phantom{\lambda}\lambda|\nu)}}^{(1)}\ .
\end{multline}
In STG,  the connection $\Gamma$ becomes symmetric because the torsion vanishes,  i.e., 
\begin{equation}\label{53}
T^{\alpha}_{\phantom{\alpha}\mu\nu}\bigl[\Gamma\bigr]=0\ .
\end{equation}
This implies at any orders
\begin{equation}
\Gamma^{\alpha}_{\phantom{\alpha}[\mu\nu]}=0\ .
\end{equation}
The absence of curvature of  connection gives
\begin{equation}\label{54}
R^{\alpha}_{\phantom{\alpha}\beta\mu\nu}\bigl[\Gamma\bigr]=0\ .
\end{equation}
It gives  additional constraints in the connection at first order, i.e.,
\begin{equation}\label{55}
{\Gamma^{\alpha}_{\phantom{\alpha}\beta\nu,\mu}}^{(1)}={\Gamma^{\alpha}_{\phantom{\alpha}\beta\mu,\nu}}^{(1)}\ .
\end{equation}
The contraction of  $\alpha$ with $\mu$ indices in Eq.~\eqref{55} gives us
\begin{equation}\label{56}
\partial_{\alpha}{\Gamma^{\alpha}_{\phantom{\alpha}\beta\nu}}^{(1)}=\partial_{\nu}{\Gamma^{\alpha}_{\phantom{\alpha}\beta\alpha}}^{(1)}\ ,
\end{equation}
and  the symmetry of  connection   in Eq.~\eqref{55} gives 
\begin{equation}\label{57}
2\partial_{\alpha}{\Gamma^{\alpha}_{\phantom{\alpha}\beta\nu}}^{(1)}=2\partial_{(\beta}{\Gamma^{\alpha}_{\phantom{\alpha}\alpha|\nu)}}^{(1)}\ .
\end{equation}
Finally,   boundary term~\eqref{16.4}, when linearized,  can be expressed as
\begin{equation}\label{58}
B^{(1)}=\Box h-\partial_{\alpha}\partial_{\beta}h^{\alpha\beta}+\partial_{\alpha}{\Gamma^{\alpha\phantom{\beta}\beta}_{\phantom{\alpha}\beta}}^{(1)}-\partial_{\alpha}{\Gamma^{\beta\phantom{\beta}\alpha}_{\phantom{\beta}\beta}}^{(1)}\ .
\end{equation}
Thanks to  Eq.~\eqref{57} for $\nu=\beta$, it reduces to
\begin{equation}\label{59}
B^{(1)}=\Box h-\partial_{\alpha}\partial_{\beta}h^{\alpha\beta}\ ,
\end{equation}
that is, at  first order in the perturbations of  metric and  connection, the boundary term is independent of the connection perturbations ${\Gamma^{\alpha}}^{(1)}_{\mu\nu}$.  Therefore, thanks to the invariance of  boundary term to the first order $B^{(1)}$ under perturbations~\eqref{40_5} and from Eq.~\eqref{57}, Eqs.~\eqref{64} and \eqref{65_4}  remain unchanged, this means that, in {\em any gauge}, the linearized field equations in vacuum are the same as those obtained in the coincidence gauge (see Ref.~\cite{CC} for  $f(Q)$ gravity).  In fact, the    metric and  connection field   Eqs.~\eqref{82} and \eqref{87}, linearized at first order, can be written in {\em any gauge} as
\begin{multline}\label{59_1}
-\frac{1}{2}f_{Q}(0,0)\Bigl[\Box h_{\mu\nu}-\bigl(\partial_{\alpha}\partial_{\mu}h^{\alpha}_{\phantom{\alpha}\nu}+\partial_{\alpha}\partial_{\nu}h^{\alpha}_{\phantom{\alpha}\mu}\bigr)-\eta_{\mu\nu}\bigl(\Box h-\partial_{\alpha}\partial_{\beta}h^{\alpha\beta}+\partial_{\alpha}{\Gamma^{\alpha\phantom{\beta}\beta}_{\phantom{\alpha}\beta}}^{(1)}-\partial_{\alpha}{\Gamma^{\beta\phantom{\beta}\alpha}_{\phantom{\beta}\beta}}^{(1)}\bigr)\\
+\partial_{\mu}\partial_{\nu}h+2\partial_{\alpha}{\Gamma^{\alpha}_{\phantom{\alpha}\mu\nu}}^{(1)}-2\partial_{(\mu}{\Gamma^{\alpha}_{\phantom{\alpha}\alpha|\nu)}}^{(1)}\Bigl]\\
-f_{BB}(0,0)\Bigl[\bigl(\partial_{\mu}\partial_{\nu}-\eta_{\mu\nu}\Box\bigr)\bigl(\Box h-\partial_{\alpha}\partial_{\beta}h^{\alpha\beta}+\partial_{\alpha}{\Gamma^{\alpha\phantom{\beta}\beta}_{\phantom{\alpha}\beta}}^{(1)}-\partial_{\alpha}{\Gamma^{\beta\phantom{\beta}\alpha}_{\phantom{\beta}\beta}}^{(1)}\bigr)\Bigr]=0\ ,
\end{multline}
and 
\begin{equation}\label{59_2}
\partial_{\mu}\partial_{\alpha}{P^{\alpha\mu}_{\phantom{\alpha\mu}\nu}}^{(1)}=0\ ,
\end{equation}
with ${P^{\alpha\mu}_{\phantom{\alpha\mu}\nu}}^{(1)}$ given by Eq.~\eqref{47_5}.  As in the coincident gauge, the first-order connection Eq.~\eqref{59_2} does not add any other constraints to those of  linearized field Eqs.~\eqref{59_1} and therefore we can ignore it.

Let us now derive  the trace of Eqs.~\eqref{59_1}. We obtain the following KG equation in {\em any gauge}
\begin{equation}\label{59_3}
\bigl(\Box+m_{B}^{2}\bigr)\bigl(\Box h-\partial_{\alpha}\partial_{\beta}h^{\alpha\beta}+\partial_{\alpha}{\Gamma^{\alpha\phantom{\beta}\beta}_{\phantom{\alpha}\beta}}^{(1)}-\partial_{\alpha}{\Gamma^{\beta\phantom{\beta}\alpha}_{\phantom{\beta}\beta}}^{(1)}\bigr)=0\ ,
\end{equation}
where $m_{B}^2=-f_{Q}(0,0)/f_{BB}(0,0)$ is  the effective mass squared  associated to the KG equation.  According to Eq.~\eqref{59_3}, Eq.~\eqref{59_1} takes the following form 
\begin{multline}\label{59_4}
-\frac{1}{2}f_{Q}(0,0)\Bigl[\Box h_{\mu\nu}-\bigl(\partial_{\alpha}\partial_{\mu}h^{\alpha}_{\phantom{\alpha}\nu}+\partial_{\alpha}\partial_{\nu}h^{\alpha}_{\phantom{\alpha}\mu}\bigr)+\partial_{\mu}\partial_{\nu}h+2\partial_{\alpha}{\Gamma^{\alpha}_{\phantom{\alpha}\mu\nu}}^{(1)}-2\partial_{(\mu}{\Gamma^{\alpha}_{\phantom{\alpha}\alpha|\nu)}}^{(1)}\\
-\frac{1}{3}\eta_{\mu\nu}\bigl(\Box h-\partial_{\alpha}\partial_{\beta}h^{\alpha\beta}+\partial_{\alpha}{\Gamma^{\alpha\phantom{\beta}\beta}_{\phantom{\alpha}\beta}}^{(1)}-\partial_{\alpha}{\Gamma^{\beta\phantom{\beta}\alpha}_{\phantom{\beta}\beta}}^{(1)}\bigr)\Bigl]\\
+f_{BB}(0,0)\Bigl[\partial_{\mu}\partial_{\nu}\bigl(\Box h-\partial_{\alpha}\partial_{\beta}h^{\alpha\beta}+\partial_{\alpha}{\Gamma^{\alpha\phantom{\beta}\beta}_{\phantom{\alpha}\beta}}^{(1)}-\partial_{\alpha}{\Gamma^{\beta\phantom{\beta}\alpha}_{\phantom{\beta}\beta}}^{(1)}\bigr)\Bigr]=0\ . 
\end{multline}
Considering  relations~\eqref{57}, \eqref{58} and \eqref{59},  Eq.~\eqref{59_4} reduces to  Eq.~\eqref{64},  which is exactly the same equation expressed in the coincidence gauge.  Furthermore, the metric field Eqs.~\eqref{64} are gauge invariant, since they do not change under infinitesimal gauge transformations.

\section{Gravitational waves}\label{DF}
Let us derive now  GW solutions  in $f(Q,B)$ gravity. Field Eqs.~\eqref{64} can be transformed in the Fourier space thanks to the following waves expansion 
\begin{equation}\label{64.5}
h_{\mu\nu}(x)=\frac{1}{(2\pi)^{3/2}}\int\,d^{3}k\;\bigl(\widetilde{h}_{\mu\nu}\bigl(\mathbf{k}\bigr)e^{ik\cdot x}+c.c.\bigr)\ ,
\end{equation}
where  $\mathbf{k}$ is the wave 3-vector. It is 
\begin{multline}\label{65}
k_{\mu}k_{\nu}\bigl(k^{2}\tilde{h}-k^{\alpha}k^{\beta}\tilde{h}_{\alpha\beta}\bigr)\\-\frac{1}{2}C\Bigl[-k^{2}\tilde{h}_{\mu\nu}+\bigl(k_{\mu}k^{\alpha}\tilde{h}_{\alpha\nu}+k_{\nu}k^{\alpha}\tilde{h}_{\alpha\mu}\bigr)+\frac{1}{3}\eta_{\mu\nu}\bigl(k^{2}\tilde{h}-k^{\alpha}k^{\beta}\tilde{h}_{\alpha\beta}\bigr)-k_{\mu}k_{\nu}\widetilde{h}\Bigr]=0\ ,
\end{multline}
where $C=f_{Q}(0,0)/f_{BB}(0,0)$ with $f_{BB}(0,0)\neq 0$. The related trace equation  is
\begin{equation}\label{66}
\bigl(k^{2}-m_{B}^{2}\bigr)\bigl(k^{2}\tilde{h}-k^{\alpha}k^{\beta}\tilde{h}_{\alpha\beta}\bigr)=0\ ,
\end{equation}
where the mass~\eqref{63_0} has been considered.
If $k^{2}\neq m_{B}^{2}$,  the trace ~\eqref{66}  further reduces to 
\begin{equation}\label{67}
k^{2}\tilde{h}-k^{\alpha}k^{\beta}\tilde{h}_{\alpha\beta}=0\ ,
\end{equation}
and  field Eqs.~\eqref{65} become
\begin{equation}\label{68}
\frac{1}{2}CF_{\mu\nu}=-\frac{1}{2}C\Bigl[-k^{2}\tilde{h}_{\mu\nu}+\bigl(k_{\mu}k^{\alpha}\tilde{h}_{\alpha\nu}+k_{\nu}k^{\alpha}\tilde{h}_{\alpha\mu}\bigr)-k_{\mu}k_{\nu}\tilde{h}\Bigr]=0\ ,
\end{equation}
which are identical to the Fourier decomposition of  linearized field equations of $f(Q)$ gravity.  This means that  if $k^{2}\neq m_{B}^{2}$,  $f(Q,B)$ gravity reproduces the same results of $f(Q)$ gravity and GWs are only massless, transverse and of spin $2$, i.e., the GR  gravitational waves (see~\cite{CCN2, CC} for details). 

Let us start our analysis  by assuming that the wave propagates along the  $+z$ direction with a wave vector $k^{\mu}=\bigl(\omega,0,0,k_{z}\bigr)$, where $k^{2}=\omega^{2}-k_{z}^{2}$. The ten components of linearized field Eqs.~\eqref{68} in  $k$-space are 
\begin{equation}\label{33}
\begin{aligned}
F_{00}&=\bigl(\omega^{2}+k_{z}^{2}\bigr)\tilde{h}_{00}+2\omega k_{z}\tilde{h}_{03}-\omega^{2}\tilde{h}=0\ ,\\
F_{01}&=k_{z}^{2}\tilde{h}_{01}+\omega k_{z}\tilde{h}_{13}=0\ ,\\
F_{02}&=k_{z}^{2}\tilde{h}_{02}+\omega k_{z}\tilde{h}_{23}=0\ ,\\
F_{03}&=\omega k_{z}\bigl(\tilde{h}_{00}-\tilde{h}_{33}-\tilde{h}\bigr)=0\ ,\\
F_{11}&=k^{2}\tilde{h}_{11}=0\ ,\\
F_{12}&=k^{2}\tilde{h}_{12}=0\ ,\\
F_{13}&=\omega k_{z}\tilde{h}_{01}+\omega^{2}\tilde{h}_{13}=0\ ,\\
F_{22}&=k^{2}\tilde{h}_{22}=0\ ,\\
F_{23}&=\omega k_{z}\tilde{h}_{02}+\omega^{2}\tilde{h}_{23}=0\ ,\\
F_{33}&=2\omega k_{z}\tilde{h}_{03}+\bigl(\omega^{2}+k_{z}^{2}\bigr)\tilde{h}_{33}+k_{z}^{2}\tilde{h}=0\ ,
\end{aligned}
\end{equation}
while the  linearized  trace Eq.~\eqref{67}, in $k$-space, becomes
\begin{equation}\label{34}
\omega^{2}\tilde{h}_{00}+2\omega k_{z}\tilde{h}_{03}+k_{z}^{2}\tilde{h}_{33}-\bigl(\omega^{2}-k_{z}^{2}\bigr)\tilde{h}=0\ ,
\end{equation}
where $\tilde{h}$ is the trace of the linear metric perturbation $h_{\mu\nu}$ in the momentum space, namely, 
\begin{equation}
\tilde{h}=\tilde{h}_{00}-\tilde{h}_{11}-\tilde{h}_{22}-\tilde{h}_{33}\ .
\end{equation}
First of all, we have to solve the set of Eqs.~\eqref{33} and  Eq.~\eqref{34} for $k^{2}\neq 0,m_{B}^{2}$ and we obtain the following solution
\begin{equation}\label{35}
\begin{aligned}
\tilde{h}_{11}&=\tilde{h}_{12}=\tilde{h}_{22}=0\ ,\\
\tilde{h}_{13}&=-\frac{k_{z}}{\omega}\tilde{h}_{01}\ ,\\
\tilde{h}_{23}&=-\frac{k_{z}}{\omega}\tilde{h}_{02}\ ,\\
\tilde{h}_{33}&=-2\frac{k_{z}}{\omega}\tilde{h}_{03}-\frac{k_{z}^{2}}{\omega^{2}}\tilde{h}_{00}\ ,
\end{aligned}
\end{equation}
with four independent variables $\tilde{h}_{01}$, $\tilde{h}_{02}$, $\tilde{h}_{03}$ and $\tilde{h}_{00}$. 
\\
Then in the case $k^{2}=0$ the solution of Eqs.~\eqref{33} and \eqref{34}, where $\omega_{1}=k_{z}$, becomes 
\begin{equation}\label{36}
\begin{aligned}
\tilde{h}_{22}&=-\tilde{h}_{11}\ ,\\
\tilde{h}_{13}&=-\tilde{h}_{01}\ ,\\
\tilde{h}_{23}&=-\tilde{h}_{02}\ ,\\
\tilde{h}_{33}&=-2\tilde{h}_{03}-\tilde{h}_{00}\ ,
\end{aligned}
\end{equation}
with six independent variables $\tilde{h}_{12}$, $\tilde{h}_{11}$, $\tilde{h}_{01}$, $\tilde{h}_{02}$, $\tilde{h}_{03}$ and $\tilde{h}_{00}$. 
\\
Finally, for $k^{2}=m^{2}_{B}$,  Eq.~\eqref{65} becomes
\begin{multline}\label{69}
H_{\mu\nu}=\frac{1}{2}m_{B}^{2}\bigl(m_{B}^{2}\eta_{\mu\nu}-k_{\mu}k_{\nu}\bigr)\tilde{h}-\bigl(\frac{1}{2}m_{B}^{2}\eta_{\mu\nu}+k_{\mu}k_{\nu}\bigr)k^{\alpha}k^{\beta}\tilde{h}_{\alpha\beta}\\
+\frac{3}{2}m_{B}^{2}\bigl(k_{\mu}k^{\alpha}\tilde{h}_{\alpha\nu}+k_{\nu}k^{\alpha}\tilde{h}_{\alpha\mu}\bigr)-\frac{3}{2}m_{B}^{4}\tilde{h}_{\mu\nu}=0\ .
\end{multline}
For a wave propagating in positive $z$ direction,  the ten components  of Eqs.~\eqref{69}, where $\omega_{2}^{2}=m_{B}^{2}+k_{z}^{2}$, are
\begin{equation}
\begin{aligned}\label{70}
H_{00}&=\bigl(m_{B}^{2}-\omega^{2}\bigr)\biggl[\frac{1}{2}m_{B}^{2}\tilde{h}+\biggl(\omega^{2}-\frac{3}{2}m_{B}^{2}\biggr)\tilde{h}_{00}+2\omega k_{z}\tilde{h}_{03}+\biggl(\frac{1}{2}m_{B}^{2}+\omega^{2}\biggr)\tilde{h}_{33}\biggr]=0\ ,\\
H_{01}&=-\frac{3}{2}m_{B}^{4}\tilde{h}_{01}=0\ ,\\
H_{02}&=-\frac{3}{2}m_{B}^{4}\tilde{h}_{02}=0\ ,\\
H_{03}&=m_{B}k_{z}\biggl[\frac{1}{2}m_{B}^{2}\tilde{h}+\omega^{2}\tilde{h}_{00}+2\omega k_{z}\tilde{h}_{03}+\biggl(\frac{3}{2}m_{B}^{2}+k_{z}^{2}\biggr)\tilde{h}_{33}\biggr]\ ,\\
H_{11}&=-\frac{1}{2}m_{B}^{4}\tilde{h}+\frac{1}{2}m_{B}^{2}\bigl(\omega^{2}\tilde{h}_{00}+2\omega k_{z}\tilde{h}_{03}+k_{z}^{2}\tilde{h}_{33}\bigr)-\frac{3}{4}m_{B}^{4}\tilde{h}_{11}=0\ ,\\
H_{12}&=-\frac{3}{2}m_{B}^{4}\tilde{h}_{12}=0\ ,\\
H_{13}&=-\frac{3}{2}m_{B}^{4}\tilde{h}_{13}=0\ ,\\
H_{22}&=-\frac{1}{2}m_{B}^{4}\tilde{h}+\frac{1}{2}m_{B}^{2}\bigl(\omega^{2}\tilde{h}_{00}+2\omega k_{z}\tilde{h}_{03}+k_{z}^{2}\tilde{h}_{33}\bigr)-\frac{3}{4}m_{B}^{4}\tilde{h}_{22}=0\ ,\\
H_{23}&=\frac{3}{2}m_{B}^{4}\tilde{h}_{23}=0\ ,\\
H_{33}&=-\omega^{2}\biggl[\frac{1}{2}m_{B}^{2}\tilde{h}+\biggl(\omega^{2}-\frac{3}{2}m_{B}^{2}\biggr)\tilde{h}_{00}+2\omega k_{z}\tilde{h}_{03}+\biggl(\frac{3}{2}m_{B}^{2}+k_{z}^{2}\biggr)\tilde{h}_{33}\biggr]=0\ .
\end{aligned}
\end{equation}
Then it is easy to get the scalar mode as solution of  system~\eqref{70} for $k^{2}=m_{B}^{2}$. It is 
\begin{equation}\label{71}
\begin{aligned}
H_{01}=H_{02}=H_{12}=H_{13}=H_{23}=0\quad&\Rightarrow\quad \tilde{h}_{01}=\tilde{h}_{02}=\tilde{h}_{12}=\tilde{h}_{13}=\tilde{h}_{23}=0\ ,\\
\frac{H_{00}}{m_{B}^{2}-\omega^{2}}+\frac{H_{33}}{\omega^{2}}=3m_{B}^{2}\tilde{h}_{33}=0\quad&\Rightarrow\quad \tilde{h}_{33}=0\ ,\\
-\frac{H_{00}}{m_{B}^{2}-\omega^{2}}+\frac{H_{03}}{m_{B}k_{z}}=\frac{3}{2}m_{B}^{2}\tilde{h}_{00}\quad&\Rightarrow\quad \tilde{h}_{00}=0\ ,\\
H_{11}-H_{22}=0\quad&\Rightarrow\quad \tilde{h}_{11}=\tilde{h}_{22}\ ,\\
\tilde{h}_{00}=\tilde{h}_{33}=0\quad \text{and}\quad H_{03}=0\quad&\Rightarrow\quad \tilde{h}_{03}=0\ ,
\end{aligned}
\end{equation}
that is, only one degree of freedom survives, $\tilde{h}_{11}$ or equivalently $\tilde{h}_{22}$.

\section{Polarizations by  the deviation equation}\label{DG}
We have now to investigate the polarizations of the above GW solutions. 
In STG with boundary term $B$,  the separation vector connecting two  free bodies traveling along nearby worldlines obeys the  geodesic deviation Eq.~\eqref{601} or \eqref{604}, due to the LC covariant conservation of the energy and momentum densities.  Therefore, since the displacement $\eta^{\mu}$ is a spacelike vector, it can be chosen as $\eta^{\mu}=(0,\boldsymbol{\chi})$ where $\boldsymbol{\chi}$ is a spatial separation vector, i.e.  $\boldsymbol{\chi}= (\chi_{x},\chi_{y},\chi_{z})$, that connects two neighboring particles with non-relativistic velocity at rest in the freely falling local frame. The spatial components of  geodesic deviation Eq.~\eqref{601} to the first order in the metric perturbation $h_{\mu\nu}$ are
\begin{equation}\label{40}
\ddot{\chi}^{i}=-\mathcal{R}^{i(1)}_{\phantom{i}0j0}\chi^{(0)j}\ ,
\end{equation}
where $i,j$ range over $(1,2,3)$, the dot above $\chi$ stands for derivative with respect to the proper time and the electric components of the linearized Riemann tensor $\mathcal{R}^{\alpha}_{\phantom{\alpha}\beta\mu\nu}$ are
\begin{equation}\label{41}
\mathcal{R}^{(1)}_{i0j0}=\frac{1}{2}\bigl(h_{i0,0j}+h_{0j,i0}-h_{ij,00}-h_{00,ij}\bigr)\ .
\end{equation}
Now we adopt a local Lorentz frame, namely a coordinate system $\{x^{\hat{\alpha}}\}$ in which the metric tensor $g_{\mu\nu}$ can be written as~\cite{CMW, MAGGIORE, MTW, CARROL} 
\begin{equation}\label{62_001}
ds^2=dx^{\hat{0}^2}-\delta_{\hat{i}\hat{j}}dx^{\hat{i}}dx^{\hat{j}}+O(\vert x^{\hat{j}}\vert^2)dx^{\hat{\alpha}}dx^{\hat{\beta}}\ ,
\end{equation}
that reduces to the Minkowski metric $\eta_{\mu\nu}$ and with its partial derivatives vanishing, in the origin of the reference system $\mathcal{P}_{0}$, that is,
\begin{equation}\label{62_01}
g_{\hat{\mu}\hat{\nu}}(\mathcal{P}_{0})=\eta_{\hat{\mu}\hat{\nu}}\quad\text{and}\quad\partial_{\hat{\alpha}}g_{\hat{\mu}\hat{\nu}}(\mathcal{P}_{0})=0\ .
\end{equation}
Eqs.~\eqref{62_01}, together with the second relation in Eq.~\eqref{40_5}, imply that, in STG, we have
\begin{equation}\label{62_02}
\Gamma^{\hat{\alpha}}_{\phantom{\alpha}\hat{\mu}\hat{\nu}}(\mathcal{P}_{0})=L^{\hat{\alpha}}_{\phantom{\alpha}\hat{\mu}\hat{\nu}}(\mathcal{P}_{0})=0\ .
\end{equation}
We choose, as local inertial frame, the proper reference frame of a freely falling test particle $\bf A$, located at the origin of frame $\mathcal{P}_{0}$.   The nearby freely falling test particle $\bf B$,  initially at rest with respect to $\bf A$, can be described by the same proper frame to lowest order, as long as it moves slowly. Then, when we take $\eta^{\hat{\alpha}}=(0,\boldsymbol{\chi)}$, the displacement 3-vector $\chi^{\hat{i}}$ represents the spatial coordinates of  particle $\bf B$  or, equivalently, its distance 3-vector by particle $\bf A$ located at origin. The comoving 4-velocities of $\bf A$ and  $\bf B$ are the same up to the first order in $h$, i.e., $u_{\bf A}^{\hat{\alpha}}=u_{\bf B}^{\hat{\alpha}}=(1,0,0,0)$ .  Then, to the first order in the linear metric perturbation $h_{\mu\nu}$,  we can identify the proper time $\tau$ with the coordinate time $t$, i.e.,
\begin{equation}\label{62_03}
t=\tau+O(h)\ .
\end{equation}
In our freely falling frame, the following identities are fulfilled,
\begin{equation}\label{62_04}
\partial_{\hat{0}}\Gamma^{\hat{i}}_{\phantom{i}\hat{0}\hat{j}}(\mathcal{P}_{0})=\partial_{\hat{0}}L^{\hat{i}}_{\phantom{i}\hat{0}\hat{j}}(\mathcal{P}_{0})=0\ ,
\end{equation}
and it follows that, up to the second order  in $h$, it holds that
\begin{equation}\label{62_05}
\frac{D^{\prime 2}\chi^{\hat{i}}}{d\tau^2}=\frac{d^{2}\chi^{\hat{i}}}{dt^2}
+O(h^2)\ .
\end{equation}
Now, from Eqs.~\eqref{62_001}--\eqref{62_05},  geodesic deviation Eq.~\eqref{604}, expressed in the STG connection, in the  locally inertial proper reference frame reduces to 
\begin{equation}\label{64_00}
\boxed{
\ddot{\chi}^{\hat{i}}=2{\partial_{[\hat{j}}L^{\hat{i}}_{\phantom{\mu}\hat{0}|\hat{0}]}}^{(1)}{\chi^{\hat{j}}}
}\ ,
\end{equation} 
where the dot stands for the derivative with respect to the coordinate time $t$.  In any gauge, from the linearized disformation tensor~\eqref{46_5},  we find 
\begin{equation}\label{63_00}
{\partial_{\nu}L^{\mu}_{\phantom{\mu}\alpha\beta}}^{(1)}=\frac{1}{2}\Bigl(\partial_{\nu}\partial^{\mu}h_{\alpha\beta}-2\partial_{\nu}\partial_{(\alpha}h^{\mu}_{\phantom{\mu}\beta)}+2\partial_{\nu}{\Gamma^{\mu}_{\phantom{\mu}\alpha\beta}}^{(1)}\Bigr)\ ,
\end{equation}
and from the constrain~\eqref{55} and Eq.~\eqref{63_00}, follows that
\begin{equation}
2{\partial_{[\nu}L^{\mu}_{\phantom{\mu}\alpha|\beta]}}^{(1)}=\frac{1}{2}\Bigl(\partial_{\alpha}\partial_{\beta}h^{\mu}_{\phantom{\mu}\nu}+\partial_{\nu}\partial^{\mu}h_{\alpha\beta}-\partial^{\mu}\partial_{\beta}h_{\alpha\nu}-\partial_{\nu}\partial_{\alpha}h^{\mu}_{\phantom{\mu}\beta}\Bigr)\ ,
\end{equation}
that is, the contribution of the linearized connection ${\Gamma^{\mu}_{\phantom{\mu}\alpha\beta}}^{(1)}$ in Eq.~\eqref{63_00} disappears.  For our components in any gauge, we get 
\begin{equation}\label{65_00}
2{\partial_{[j}L^{i}_{\phantom{\mu}0|0]}}^{(1)}=\frac{1}{2}\Bigl(\partial_{0}\partial_{0}h^{i}_{\phantom{\mu}j}+\partial_{j}\partial^{i}h_{00}-\partial^{i}\partial_{0}h_{0j}-\partial_{j}\partial_{0}h^{i}_{\phantom{\mu}0}\Bigr)\ .
\end{equation}
From the gauge invariance of  Eq.~\eqref{65_00}, we have 
\begin{equation}
2{\partial_{[j}L^{i}_{\phantom{\mu}0|0]}}^{(1)}=2{\partial_{[\hat{j}}L^{\hat{i}}_{\phantom{\mu}\hat{0}|\hat{0}]}}^{(1)}\ ,
\end{equation}
which allows us to recast Eq.~\eqref{64_00} into the following form
\begin{equation}\label{66_0}
\boxed{
\ddot{\chi}^{\hat{i}}=2{\partial_{[j}L^{i}_{\phantom{\mu}0|0]}}^{(1)}{\chi^{\hat{j}}}
}\ .
\end{equation} 
It is  the geodesic deviation equation for $f(Q,B)$ gravity, in the proper reference frame of freely falling particles where ${\partial_{[j}L^{i}_{\phantom{\mu}0|0]}}^{(1)}$ is expressed in {\em any gauge}.   Since   in Eq.~\eqref{65_00}.  Eq.~\eqref{66_0} there are the second derivatives of metric perturbations  $h_{\mu\nu}$, it can be regarded as the relative acceleration between  two freely falling point particles. 

Inserting  Eq.~\eqref{65_00} in Eq.~\eqref{66_0}, the linear system of differential equations for a wave traveling along positive $z$-axis in the local proper reference reads as
\begin{equation}\label{42}
\left\{
\begin{array}{lr}
\ddot{\chi}_{x}=-\frac{1}{2}h_{11,00}\chi_{x}^{0}-\frac{1}{2}h_{12,00}\chi_{y}^{0}+\frac{1}{2}\bigl(h_{01,03}-h_{13,00}\bigr)\chi_{z}^{0}\\
\ddot{\chi}_{y}=-\frac{1}{2}h_{12,00}\chi_{x}^{0}-\frac{1}{2}h_{22,00}\chi_{y}^{0}+\frac{1}{2}\bigl(h_{02,03}-h_{23,00}\bigr)\chi_{z}^{0}\\
\ddot{\chi}_{z}=\frac{1}{2}\bigl(h_{01,03}-h_{13,00}\bigr)\chi_{x}^{0}+\frac{1}{2}\bigl(h_{02,03}-h_{23,00}\bigr)\chi_{y}^{0}+\frac{1}{2}\bigl(2h_{03,03}-h_{33,00}-h_{00,33}\bigr)\chi_{z}^{0}
\end{array}
\right.\ ,
\end{equation}
according to ${\partial_{j}L^{i}_{\phantom{i}00}}^{(1)}=-{\partial_{j}L_{i00}}^{(1)}$ in linear approximation with $\eta_{\mu\nu}=\text{diag}(1,-1,-1,-1)$ and where $\boldsymbol{\chi}^{0}$ stands for the initial spatial position of particle $\bf B$ while $\boldsymbol{\chi}(t)$ stands for its spatial position at the coordinate time $t$ when the wave hits it.

In the case of non-null wave $k^{2}\neq 0,m_{B}^{2}$,  from the solution~\eqref{35} for the following wave expansion
\begin{equation}\label{43}
h_{\mu\nu}(z,t)=\frac{1}{\sqrt{2\pi}}\int dk_{z}\Bigl(\tilde{h}_{\mu\nu}(k_{z})e^{i(\omega t-k_{z}z)}+c.c.\Bigr)\ ,
\end{equation}
where $c.c.$ stands for complex conjugate, the system~\eqref{42} takes the form
\begin{equation}\label{44}
\left\{
\begin{array}{lr}
\ddot{\chi}_{x}=0\\
\ddot{\chi}_{y}=0\\
\ddot{\chi}_{z}=0
\end{array}
\right.\ .
\end{equation}
Then, imposing the initial conditions $\boldsymbol{\chi}(0)=\mathbf{R}_{0}=(\chi_{x}^{0},\chi_{y}^{0},\chi_{z}^{0})$ and $\dot{\boldsymbol{\chi}}(0)={\bf 0}$,  after double integration with respect to the coordinate time $t$,  the system~\eqref{44} gives the following solution
\begin{equation}\label{45}
\chi_{x}(t)=\chi_{x}^{0},\quad \chi_{y}(t)=\chi_{y}^{0},\quad \chi_{z}(t)=\chi_{z}^{0}\ ,
\end{equation}
that is, there is no mode associated with $k^{2}\neq 0$ or $k^{2}\neq m_{B}^{2}$ .
\\
\\
On the other hand, in the case of null wave $k^{2}=0$,  for a plane wave at $k_{z}$  with angular frequency $\omega_{1}=k_{z}$ propagating along $+z$-direction, by means of  solution~\eqref{36} and  expansion~\eqref{43},  the equation of geodesic deviation at first order in the displacement~\eqref{42} yields
\begin{equation}\label{46}
\left\{
\begin{array}{lr}
\ddot{\chi}_{x}=\frac{1}{2}\omega_{1}^{2}\bigl(\tilde{h}^{(+)}\chi_{x}^{0}+\tilde{h}^{(\times)}\chi_{y}^{0}\bigr)e^{i\omega_{1}(t-z)}\\
\ddot{\chi}_{y}=\frac{1}{2}\omega_{1}^{2}\bigl(\tilde{h}^{(\times)}\chi_{x}^{0}-\tilde{h}^{(+)}\chi_{y}^{0}\bigr)e^{i\omega_{1}(t-z)}\\
\ddot{\chi}_{z}=0
\end{array}
\right.+ c.c.\ ,
\end{equation}
where $\tilde{h}_{11}=-\tilde{h}_{22}=\tilde{h}^{(+)}$ and $\tilde{h}_{12}=\tilde{h}^{(\times)}$. Then after double integration with respect to $t$ the solution of system~\eqref{46} becomes 
\begin{equation}
\left\{
\begin{array}{lr}
\chi_{x}(t)=\chi_{x}^{0}-\frac{1}{2}\bigl(\tilde{h}^{(+)}\chi_{x}^{0}+\tilde{h}^{(\times)}\chi_{y}^{0}\bigr)e^{i\omega_{1}(t-z)}\\
\chi_{y}(t)=\chi_{y}^{0}-\frac{1}{2}\bigl(\tilde{h}^{(\times)}\chi_{x}^{0}-\tilde{h}^{(+)}\chi_{y}^{0}\bigr)e^{i\omega_{1}(t-z)}\\
\chi_{z}(t)=\chi_{z}^{0}
\end{array}
\right.+ c.c.\ ,
\end{equation}
describing the well-known  massless,   spin-2 transverse, plus and cross modes  of GR.
\\\\
Finally, the  system of  Eqs.~\eqref{42} for the deviation vector $\boldsymbol{\chi}$,  describing an elementary massive wave at a given $k_{z}$, with $k^{2}=m_{B}^{2}=-f_{Q}(0,0)/3f_{BB}(0,0)$, associated to the KG equation for linearized $B^{(1)}$~\eqref{62_KG} and propagating along the $+z$ axis according to the solution~\eqref{71} and the expansion~\eqref{43},  yields 
\begin{equation}\label{80}
\left\{
\begin{array}{lr}
\ddot{\chi}_{x}=\frac{1}{2}\omega_{2}^{2}\tilde{h}^{(s)}\chi_{x}^{0}\,e^{i(\omega_{2} t-k_{z}z)}\\
\ddot{\chi}_{y}=\frac{1}{2}\omega_{2}^{2}\tilde{h}^{(s)}\chi_{y}^{0}\,e^{i(\omega_{2} t-k_{z}z)}\\
\ddot{\chi}_{z}=0
\end{array}
\right.+c.c.\ ,
\end{equation}
where $\tilde{h}^{(s)}=\tilde{h}_{11}=\tilde{h}_{22}$ with $|\tilde{h}^{(s)}|$ very small. When we are sufficiently far from the  radiation source, we can suppose that $m_{B}^2$ is very small, i.e., that the plane waves are nearly null. Then, we can expand our quantities with respect to a parameter $\gamma$, which takes into account the difference in speed between nearly null plane waves with speed $v_{g}$ and exactly null plane waves with speed $c$ where  $\gamma$ parameter vanishes. Therefore, the expansion coefficient $\gamma$ reads
~\cite{CMW} 
\begin{equation}\label{310}
\gamma=\left(\frac{c}{v_{g}}\right)^{2}-1\ .
\end{equation} 
The $\gamma$ parameter can be rewritten as 
\begin{equation}\label{311}
\frac{m_{B}^{2}}{k_{z}^{2}}=\frac{1}{k_{z}^{2}}\left[\frac{\omega^{2}}{c^{2}}-k_{z}^{2}\right]=\left(\frac{\omega}{c k_{z}}\right)^{2}-1=\gamma\ .
\end{equation}
By using the Landau symbols,  namely little-$o$ and big-$\mathcal{O}$ notation,  we can keep $k_{z}$ fixed and we can expand in terms of our parameter $\gamma$ the following quantities in $c=1$ units, i.e.,
\begin{equation}\label{312}
\frac{\omega}{k_{z}}=\sqrt{1+\gamma}=\left(1+\frac{1}{2}\gamma\right)+o\left(\gamma\right)\ ,
\end{equation}
\begin{equation}\label{313}
e^{i\left(\omega t-k_{z}z\right)}=e^{ik_{z}\left(t-z\right)}+\mathcal{O}\left(\gamma\right)\ ,
\end{equation}
from which the first order in $\gamma$ is
\begin{equation}\label{314}
\frac{m_{B}^{2}}{\omega^{2}}=\gamma+o\left(\gamma\right)\ .
\end{equation}
After double integration with respect to $t$ of  system~\eqref{80}, according to the previous expansions,  we obtain the solution at first order in the $\gamma$ parameter, that is
\begin{equation}\label{81}
\left\{
\begin{array}{lr}
\chi_{x}(t)=\chi_{x}^{0}-\frac{1}{2}\chi_{x}^{0}\tilde{h}^{(s)}\,e^{i\omega_{2} (t-z)}+\mathcal{O}\left(\gamma\right)\\
\chi_{y}(t)=\chi_{y}^{0}-\frac{1}{2}\chi_{y}^{0}\tilde{h}^{(s)}\,e^{i\omega_{2} (t-z)}+\mathcal{O}\left(\gamma\right)\\
\chi_{z}(t)=\chi_{z}^{0}+\mathcal{O}\left(\gamma\right)
\end{array}
\right.
+ c.c.\ ,
\end{equation}
The variation $\Delta \chi_{z}=\chi_{z}(t)-\chi_{z}^{0}$ is a  higher order infinitesimal than $\Delta \chi_{x}=\chi_{x}(t)-\chi_{x}^{0}$ or $\Delta \chi_{y}=\chi_{y}(t)-\chi_{y}^{0}$ when $\gamma$ tends to zero, i.e.,
\begin{equation}
\Delta \chi_{z}=o(\Delta \chi_{x})=o(\Delta \chi_{y})\quad\text{for}\quad\gamma\rightarrow 0\ .
\end{equation}
Therefore, GWs in $f(Q,B)$ non-metric gravity oscillate to lower order in the $xy$ plane orthogonal to its direction of propagation $z$,  and then they are transversely polarized. 
So Eq.~\eqref{81} describes the response of two free falling nearby point-like particles when struck by a GW with scalar transverse mode of mass $m_{B}$,  angular frequency $\omega_{2}=\sqrt{m_{B}^{2}+k_{z}^{2}}$ and group velocity $v_{G}$ of the associated wave packet, i.e. 
\begin{equation}
v_{G}(k_{z})=\frac{\partial\omega_{2}(k_{z})}{\partial k_{z}}=\frac{k_{z}}{\omega_{2}}< c\ .
\end{equation} 
From Eq.~\eqref{81},  we explicitly obtain the wave solution
\begin{equation}
\left\{
\begin{array}{lr}
\chi_{x}(t)=\chi_{x}^{0}\Bigl[1-|\tilde{h}^{(s)}| \cos\bigl[\omega_{2}(t-z)+\beta(k_{z})\bigr]\Bigr]\\
\chi_{y}(t)=\chi_{y}^{0}\Bigl[1-|\tilde{h}^{(s)}| \cos\bigl[\omega_{2}(t-z)+\beta(k_{z})\bigr]\Bigr]\\
\chi_{z}(t)=\chi_{z}^{0}
\end{array}
\right.\ ,
\end{equation}
and then we get the circle equation of radius $r$ in the $xy$ plane
\begin{equation}
\chi_{x}^{2}+\chi_{y}^2=r^2\ ,
\end{equation}
where 
\begin{equation}
r=r_{0}\rho\quad\text{with}\quad r_{0}^{2}={\chi_{x}^{0}}^2+{\chi_{y}^{0}}^{2}\quad\text{and}\quad\rho(t)=1-|\tilde{h}^{(s)}| \cos\bigl[\omega_{2}(t-z)+\beta(k_{z})\bigr]\ .
\end{equation}
Therefore, a circle of particles of radius $r_{0}$,   placed orthogonal to the direction of wave propagation,  will oscillate between the two circles of minimum and maximum radius,  $r_{0}(1-|\tilde{h}^{s})|\leq r\leq r_{0}(1+|\tilde{h}^{s}|)$, when hit by a GW.
An arbitrary gravitational wave propagating along $+z$ direction, with massless mode $k_{1}^{2}=0$ and massive mode $k_{2}^{2}=m_{B}^{2}$,  with  wave vector $k_{1}^{\mu}=( \omega_{1},0,0,k_{z})$ and $k_{2}^{\mu}=( \omega_{2},0,0,k_{z})$ respectively,  can be expressed in terms of the three polarization tensors $\epsilon_{\mu\nu}^{(+)}$, $\epsilon_{\mu\nu}^{(\times)}$ and $\epsilon_{\mu\nu}^{(s)}$ as
\begin{multline}\label{305}
h_{\mu\nu}\bigl(t,z\bigr)=\int \frac{d\omega_{1}}{\sqrt{\pi}}\,\Bigl[\tilde{h}^{(+)}\bigl(\omega_{1}\bigr)\epsilon_{\mu\nu}^{(+)}+\tilde{h}^{(\times)}\bigl(\omega_{1}\bigr)\epsilon_{\mu\nu}^{(\times)}\Bigr]e^{i\omega_{1}\bigl(t-z\bigr)}\\+\int \frac{dk_{z}}{\sqrt{\pi}}\,\tilde{h}^{(s)}\bigl(k_{z}\bigr)\epsilon_{\mu\nu}^{(s)}e^{i(\omega_{2}t-k_{z}z)}+ c.c.\ ,
\end{multline}
where 
\begin{align}\label{306}
\epsilon^{(+)}_{\mu\nu}&=\frac{1}{\sqrt{2}}
\begin{pmatrix} 
0 & 0 & 0 & 0 \\
0 & 1 & 0 & 0 \\
0 & 0 & -1 & 0 \\
0 & 0 & 0 & 0
\end{pmatrix}\ , &
\epsilon^{(\times)}_{\mu\nu}&=\frac{1}{\sqrt{2}}
\begin{pmatrix} 
0 & 0 & 0 & 0 \\
0 & 0 & 1 & 0 \\
0 & 1 & 0 & 0 \\
0 & 0 & 0 & 0
\end{pmatrix}\ , &
\epsilon^{(s)}_{\mu\nu}&=\frac{1}{\sqrt{2}}
\begin{pmatrix} 
0 & 0 & 0 & 0 \\
0 & 1 & 0 & 0 \\
0 & 0 & 1 & 0 \\
0 & 0 & 0 & 0
\end{pmatrix}\ .
\end{align}
The set of polarization tensors $\left\{\epsilon_{\mu\nu}^{\left(+\right)}, \epsilon_{\mu\nu}^{\left(\times\right)}, \epsilon_{\mu\nu}^{\left(s\right)}\right\}$ fulfills the orthonormality relations
\begin{equation}\label{306_5}
\text{Tr}\left\{\epsilon^{(a)}\epsilon^{(b)}\right\}=\epsilon_{\mu\nu}^{(a)}\epsilon^{(b)\mu\nu}=\delta^{ab}\quad\text{with}\quad a,b\in\left\{+, \times, s\right\}\ .
\end{equation} 
The helicity of wave Eq.~\eqref{305}, for any single mode, can be derived by performing a rotation of $\theta$ angle around the $z$ propagation direction, on the polarization tensors as follows
\begin{equation}\label{306_6}
\epsilon_{\mu\nu}^{\prime}=R_{\mu}^{\alpha}\epsilon_{\alpha\beta}R_{\beta}^{\nu}\ ,
\end{equation}
where the matrix of rotation is
\begin{equation}
R=
\begin{pmatrix}
1	&	0	&	0	&	0\\
0	&	\cos\theta & \sin\theta	&	0\\
0	&	-\sin\theta	&	\cos\theta & 0\\
0	&	0	&	0	&	1
\end{pmatrix}\ .
\end{equation}
Hence the wave has helicity $s$, if the polarization tensor changes as
\begin{equation}
\epsilon_{\mu\nu}^{\prime}=e^{\pm i s\theta}\epsilon_{\mu\nu}\ .
\end{equation}
Thus, for the $\omega_{1}$ mode with $k_{1}^{2}=0$, we consider a suitable linear combination of   polarization tensors $\epsilon_{\mu\nu}^{(+)}$ and $\epsilon_{\mu\nu}^{(\times)}$ with complex coefficients,  right and left circular polarizations as
\begin{equation}
\epsilon_{\mu\nu}^{R}=\frac{1}{2}\Bigl[\epsilon_{\mu\nu}^{(+)}+i\epsilon_{\mu\nu}^{(\times)}\Bigr]\ ,
\end{equation}
and 
\begin{equation}
\epsilon_{\mu\nu}^{L}=\frac{1}{2}\Bigl[\epsilon_{\mu\nu}^{(+)}-i\epsilon_{\mu\nu}^{(\times)}\Bigr]\ .
\end{equation}
Then we perform the rotation~\eqref{306_6} and $\epsilon_{\mu\nu}^{R}$, $\epsilon_{\mu\nu}^{L}$ circular polarization tensors transform as
\begin{equation}
\epsilon^{R/L\prime}_{\mu\nu}=e^{\pm 2i\theta}\epsilon^{R/L}_{\mu\nu}\ ,
\end{equation}
that is, we have obtained that  modes $\omega_{1}$ have helicity equal to two, i.e.  they are tensor modes. While for $\omega_{2}$ mode with $k_{2}^{2}=m_{B}^{2}\neq 0$, the polarization tensor $\epsilon_{\mu\nu}^{(s)}$ transforms as 
\begin{equation}
\epsilon^{(s)\prime}_{\mu\nu}=\epsilon^{(s)}_{\mu\nu}\ ,
\end{equation}
that is, we have obtained that the $\omega_{2}$ mode has helicity equal to zero, i.e. it is a scalar mode.

\section{Discussion and Conclusions}\label{E}
In this paper, we discussed the existence, nature and properties of the polarization modes of GWs in $f(Q,B)$ non-metric gravity both in coincident gauge,  turning off the STG connection, and in free gauge, without gauge fixing.  The main result is that,  regardless of the gauge, the boundary term $B$ induces a further scalar mode of GWs more than the two standard $(+)$ and $(x)$ of GR.  Unlike $f(Q)$ theories, the presence of $B$ excites an additional scalar polarization, transverse and massive with helicity zero,   group velocity $v_{G}$ less than $c$.  In other words, three degrees of freedom propagate considering  the linearized  $f(Q,B)$ theory: with amplitudes $\tilde{h}^{(+)}$ and $\tilde{h}^{(\times)}$  for tensor perturbations and amplitude $\tilde{h}^{(s)}$  for the scalar perturbation.  

Specifically, we first found an expression of the energy and momentum balance equations for generic theories of gravity which do not satisfy the LC covariant conservation of the energy-momentum tensor of a perfect fluid as well as the non-geodesic equation of motion and the deviation equation.  In general a free particle follows a non-geodesic motion but in $f(Q,B)$ gravity, as in $f(Q)$ gravity, it follows metric geodesics because the matter stress-energy tensor is LC covariantly conserved on-shell, i.e. $\mathcal{D}_{\alpha}T^{\alpha}_{\phantom{\alpha}\beta}=0$.  So, even if the test mass does not follow the autoparallel curves with respect to the STG connection but timelike geodesics, the separation vector connecting two neighboring worldlines obeys the geodesic deviation equation exactly as in GR.  In order to probe  this statement,  we derived the field and connection equations from a variational principle, keeping metric and connection independent, according to the Palatini approach.  Thus, we perturbed the metric tensor and the STG connection around the flat spacetime and, due to the independence of the linearized boundary term $B^{(1)}$ from the linearized STG connection ${\Gamma^{\alpha}}^{(1)}_{\mu\nu}$, the linearized metric and connection field equations in vacuum appear exactly the same in both the coincidence gauge and the free gauge.  In the Fourier space, we have solved the linearly perturbed field equations by looking for wavelike solutions, in three cases: $k^2\neq 0,m_{B}^2$, $k^2=0$ and $k^2=m_{B}^2$, where  $m_{B}$ is the effective mass derived  from the  Klein-Gordon equation for the  $B^{(1)}$.  Finally, by the geodesic deviation equation, we studied the polarization and helicity of GWs.  In $f(Q,B)$ non-metric gravity, we found massless transverse tensor modes with helicity equal to $2$ for null gravitational waves with $k^2=0$ as in GR and in $f(Q)$ non-metric gravity.  Furthermore, thanks to the boundary term $B$, our model exhibits a further massive transverse scalar mode with helicity $0$ for nearly null GWs with $k^{2}=m_{B}^{2}$ and mass equal to $m_ {B}$. It was assumed small enough to allow the spatial separation vector $\boldsymbol{\chi}$ to be expanded with respect to a parameter $\gamma$ which takes into account the difference in speed between the nearly null and null plane waves. Hence the GWs of $f(Q,B)$ non-metric gravity behaves as $f(R)$ metric gravity: in  both cases a massive transverse scalar mode appear due to the fact that both theories can be reduced to GR plus a scalar field \cite{CCDL}.  

Experimentally, GR and $f(T)$ models~\cite{Aldrovandi,Bamba,Ong:2013qja, Izumi:2012qj} cannot be distinguished from $f(Q)$ gravity via GW measurements because they show the same tensor polarization modes. Similarly,  $f(R)$ gravity also cannot be distinguished from  $f(Q,B)$  via GW measurements because they show the same scalar polarization mode~\cite{Caprep}. The linear gravitational actions of curvature scalar $\mathcal{R}$,  torsion scalar $T$ and non-metricity scalar  $Q$ are equivalent, in the sense that all three theories are governed by the same field equations, i.e. the well-known geometric trinity of gravity,  while their non-linear extensions $f(R)$, $f(T)$, and $f(Q)$ are not. This non-equivalence involves new physical degrees of freedom that can help us to find  possible signature of a viable cosmological model~\cite{Koussour,Mandal,Naik,Vishwakarma,Shi,Ferreira,Bajardi,Koussour1}.  The analogy between  $f(Q,B)$ and $f(R)$ gravities, due to the introduction of the boundary term $B$, is much deeper than their analogous behavior via gravitational waves. As discussed in \cite{Carmen}, in any extended theory of gravity, the boundary terms have a main role as further degrees of freedom giving rise to GW scalar modes. 
As discussed in   \cite{Takeda},  GW scalar polarization modes  are  introduced to study  viable extensions of gravity. However,   scalar mode couplings to the standard  matter are strongly constrained by several precision  experiments. This means that  the amplitude of any scalar polarization in the observed GW signals can be   significantly suppressed if compared to   tensor modes \cite{NojiriGW}. This means that sensitivity of forthcoming experiments should be highly improved to have chances to detect such modes. Very likely, the forthcoming LISA \cite{LISA} and Einstein Telescope \cite{ET} will give a final answer to this problem.

\appendix
\section{Appendix}\label{F}
\subsection{Relations between  non-metricity and disformation tensor }\label{F_0}
We show some relations connecting the non-metricity tensor $Q_{\alpha\mu\nu}$, the metric tensor $g_{\mu\nu}$ and the disformation tensor $L^{\alpha}_{\phantom{\alpha}\mu\nu}$, namely,
\begin{align}
Q^{\alpha\mu\nu}&=-\nabla^{\alpha}g^{\mu\nu}\ ,\\
Q_{\alpha\phantom{\mu}\nu}^{\phantom{\alpha}\mu}&=-g_{\rho\nu}\nabla_{\alpha}g^{\mu\rho}\ ,\\
Q_{\alpha\mu}^{\phantom{\alpha\mu}\nu}&=-g_{\mu\rho}\nabla_{\alpha}g^{\nu\rho}\ ,\\
Q_{\alpha}^{\phantom{\alpha}\mu\nu}&=-\nabla_{\alpha}g^{\mu\nu}\ ,\\
Q^{\alpha}_{\phantom{\alpha}\mu\nu}&=\nabla^{\alpha}g_{\mu\nu}\ ,
\end{align}
\begin{align}
L^{\lambda}_{\phantom{\lambda}\alpha\lambda}&=-\frac{1}{2}Q_{\alpha}\label{8.6}\ ,\\
L^{\alpha\lambda}_{\phantom{\alpha\lambda}\lambda}&=\frac{1}{2}Q^{\alpha}-\widetilde{Q}^{\alpha}\ .\label{8.7}
\end{align}
\subsection{Covariant derivatives}\label{F_1}
We explicitly write the STG covariant derivatives $\nabla_{\mu}$ of some geometric objects
\begin{align}
\nabla_{\mu}A_{\nu}&=\mathcal{D}_{\mu}A_{\nu}-L^{\alpha}_{\phantom{\alpha}\nu\mu}A_{\alpha}=\mathcal{D}_{\mu}A_{\nu}-\frac{1}{2}Q^{\alpha}_{\phantom{\alpha}\mu\nu}A_{\alpha}+Q_{(\mu\phantom{\alpha}\nu)}^{\phantom{(\mu}\alpha}A_{\alpha}\ ,\\
\nabla_{\mu}A^{\nu}&=\mathcal{D}_{\mu}A^{\nu}+L^{\nu}_{\phantom{\nu}\lambda\mu}A^{\lambda}\ ,\\
\nabla_{\mu}A_{\alpha\beta}&=\mathcal{D}_{\mu}A_{\alpha\beta}-L^{\lambda}_{\phantom{\lambda}\alpha\mu}A_{\lambda\beta}-L^{\lambda}_{\phantom{\lambda}\beta\mu}A_{\alpha\lambda}\ ,\\
\nabla_{\mu}A^{\alpha\beta}&=\mathcal{D}_{\mu}A^{\alpha\beta}+L^{\alpha}_{\phantom{\alpha}\lambda\mu}A^{\lambda\beta}+L^{\beta}_{\phantom{\beta}\lambda\mu}A^{\alpha\lambda}\ ,\\
\nabla_{\mu}A^{\alpha}_{\phantom{\alpha}\beta}&=\mathcal{D}_{\mu}A^{\alpha}_{\phantom{\alpha}\beta}+L^{\alpha}_{\phantom{\alpha}\lambda\mu}A^{\lambda}_{\phantom{\lambda}\beta}-L^{\lambda}_{\phantom{\lambda}\mu\beta}A^{\alpha}_{\phantom{\alpha}\lambda}\ ,\\
\nabla^{\alpha}A_{\alpha}&=\mathcal{D}^{\alpha}A_{\alpha}-\frac{1}{2}Q^{\alpha}A_{\alpha}+\widetilde{Q}^{\alpha}A_{\alpha}\ ,\\
&=\mathcal{D}^{\alpha}A_{\alpha}-L^{\alpha\lambda}_{\phantom{\alpha\lambda}\lambda}A_{\alpha}\\
\nabla_{\alpha}A^{\alpha}&=\nabla^{\alpha}A_{\alpha}-\widetilde{Q}^{\alpha}A_{\alpha}=\mathcal{D}_{\alpha}A^{\alpha}+L^{\lambda}_{\phantom{\lambda}\alpha\lambda}A^{\alpha}\ ,\\
 &=\mathcal{D}_{\alpha}A^{\alpha}-\frac{1}{2}Q^{\alpha}A_{\alpha}\ ,\\
\nabla_{\alpha}A^{\alpha}_{\phantom{\alpha}\beta}&=\mathcal{D}_{\alpha}A^{\alpha}_{\phantom{\alpha}\beta}-\frac{1}{2}Q_{\lambda}A^{\lambda}_{\phantom{\lambda}\beta}-L^{\lambda}_{\phantom{\lambda}\alpha\beta}A^{\alpha}_{\phantom{\alpha}\lambda}\ ,\label{8.17}\\
\nabla_{\beta}F_{\mu\nu\alpha}=&\,\mathcal{D}_{\beta}F_{\mu\nu\alpha}-L^{\lambda}_{\phantom{\lambda}\beta\mu}F_{\lambda\nu\alpha}-L^{\lambda}_{\phantom{\lambda}\beta\nu}F_{\mu\lambda\alpha}-L^{\lambda}_{\phantom{\lambda}\beta\alpha}F_{\mu\nu\lambda\ ,}\label{8.18}
\end{align}
\begin{equation}
\bigl[\nabla_{\mu},\nabla_{\nu}\bigr]T^{\mu}_{\phantom{\mu}\nu}=0\ .
\end{equation}
The non-metricity tensor $Q_{\alpha\mu\nu}$ satisfies the following Bianchi identity
\begin{equation}\label{8.8}
\nabla_{[\alpha}Q_{\beta]\mu\nu}=0\ .
\end{equation}
\subsection{Tensor densities}\label{F_2}
We define the STG covariant derivative of the $(1,1)$-tensor density $\mathcal{B}^{\mu}_{\phantom{\mu}\nu}$ of weight $w_{1}$ as
\begin{equation}
\nabla_{\alpha}\mathcal{B}^{\mu}_{\phantom{\mu}\nu}=\partial_{\alpha}\mathcal{B}^{\mu}_{\phantom{\mu}\nu}+\Gamma^{\mu}_{\phantom{\mu}\alpha\lambda}\mathcal{B}^{\lambda}_{\phantom{\lambda}\nu}-\Gamma^{\lambda}_{\phantom{\lambda}\alpha\nu}\mathcal{B}^{\mu}_{\phantom{\mu}\lambda}-w_{1}\Gamma^{\lambda}_{\phantom{\lambda}\lambda\alpha}\mathcal{B}^{\mu}_{\phantom{\mu}\nu}\ .
\end{equation}
The STG covariant divergence of a (2,1)-tensor density of weight one $\mathcal{M}^{\alpha\mu}_{\phantom{\alpha\mu}\nu}$,  leads to
\begin{equation}
\nabla_{\alpha}\mathcal{M}^{\alpha\mu}_{\phantom{\alpha\mu}\nu}=\mathcal{D}_{\alpha}\mathcal{M}^{\alpha\mu}_{\phantom{\alpha\mu}\nu}+L^{\mu}_{\phantom{\mu}\alpha\lambda}\mathcal{M}^{\alpha\lambda}_{\phantom{\alpha\lambda}\nu}-L^{\lambda}_{\phantom{\lambda}\alpha\nu}\mathcal{M}^{\alpha\mu}_{\phantom{\alpha\mu}\lambda}\ .
\end{equation}
The STG covariant divergence of a (1,1)-tensor density of weight one $\mathcal{E}^{\alpha}_{\phantom{\alpha}\nu}$, leads to 
\begin{equation}\label{8.21}
\nabla_{\alpha}\mathcal{E}^{\alpha}_{\phantom{\alpha}\nu}=\mathcal{D}_{\alpha}\mathcal{E}^{\alpha}_{\phantom{\alpha}\nu}-L^{\lambda}_{\phantom{\lambda}\alpha\nu}\mathcal{E}^{\alpha}_{\phantom{\alpha}\lambda}\ .
\end{equation}
While the STG covariant divergence of the vector density $\mathcal{C}^{\alpha}$ of weight $w$, it gets for the symmetry of the STG connection 
\begin{equation}
\nabla_{\alpha}\mathcal{C}^{\alpha}=\partial_{\alpha}\mathcal{C}^{\alpha}+\Gamma^{\alpha}_{\phantom{\alpha}\alpha\lambda}\mathcal{C}^{\lambda}-w\Gamma^{\alpha}_{\phantom{\alpha}\alpha\lambda}\mathcal{C}^{\lambda}\ .
\end{equation}
If the vector density $\mathcal{C}^{\alpha}$ is of weight $w=1$, we have
\begin{equation}
\nabla_{\alpha}\mathcal{C}^{\alpha}=\partial_{\alpha}\mathcal{C}^{\alpha}\ .
\end{equation}
For the scalar density $\sqrt{-g}$ of weight $w=1$ and for the contravariant and covariant vector densities $\mathcal{A}_{\nu}=\sqrt{-g}A_{\nu}$,, $\mathcal{A}^{\alpha}=\sqrt{-g}A^{\alpha}$, $\mathcal{A}_{\alpha}=\sqrt{-g}A_{\alpha}$ of weight $w=1$,  and the $(0,3)$-tensor density $\mathcal{F}_{\mu\nu\alpha}=\sqrt{-g}F_{\mu\nu\alpha}$ of weight $w=1$, we find
\begin{align}
\nabla_{\alpha}\sqrt{-g}=&\,\frac{1}{2}\sqrt{-g}Q_{\alpha}\ ,\label{8.24}\\
\nabla_{\mu}\bigl(\sqrt{-g}A_{\nu}\bigr)=&\,\sqrt{-g}\mathcal{D}_{\mu}A_{\nu}-\sqrt{-g}L^{\lambda}_{\phantom{\lambda}\mu\nu}A_{\lambda}+\frac{1}{2}\sqrt{-g}\,Q_{\mu}A_{\nu}\ ,\\
\nabla_{\alpha}\bigl(\sqrt{-g}A^{\alpha}\bigr)=&\,\mathcal{D}_{\alpha}\bigl(\sqrt{-g}A^{\alpha}\bigr)=\sqrt{-g}\mathcal{D}_{\alpha}A^{\alpha}\ ,\\
\nabla^{\alpha}\bigl(\sqrt{-g}A_{\alpha}\bigr)=&\,\mathcal{D}_{\alpha}\bigl(\sqrt{-g}A^{\alpha}\bigr)+\sqrt{-g}\widetilde{Q}^{\alpha}A_{\alpha}\ ,\\
\nabla_{\beta}\bigl(\sqrt{-g}F_{\mu\nu\alpha}\bigr)=&\,\sqrt{-g}\mathcal{D}_{\beta}F_{\mu\nu\alpha}-\sqrt{-g}L^{\lambda}_{\phantom{\lambda}\beta\mu}F_{\lambda\nu\alpha}-\sqrt{-g}L^{\lambda}_{\phantom{\lambda}\beta\nu}F_{\mu\lambda\alpha}\nonumber\\
&-\sqrt{-g}L^{\lambda}_{\phantom{\lambda}\beta\alpha}F_{\mu\nu\lambda}+\frac{1}{2}\sqrt{-g}Q_{\beta}F_{\mu\nu\alpha}\ .
\end{align}
\subsection{Integrals and Stokes's theorem}\label{F_3}
We explicitly display Stoke's theorem and useful integrals
\begin{align}
\int_{\Omega}\nabla_{\alpha}\bigl(\sqrt{-g}A^{\alpha}\bigr)d^{4}x&=\int_{\Omega}d^{4}x\sqrt{-g}\mathcal{D}_{\alpha}A^{\alpha}=\int_{\partial\Omega}dS_{\alpha}\sqrt{-g}A^{\alpha}\ ,\\
\int_{\Omega}\nabla^{\alpha}\bigl(\sqrt{-g}A_{\alpha}\bigr)d^{4}x&=\int_{\Omega}d^{4}x\sqrt{-g}\mathcal{D}_{\alpha}A^{\alpha}+\int_{\Omega}d^{4}x\sqrt{-g}\,\widetilde{Q}^{\alpha}A_{\alpha}\nonumber\\
&=\int_{\partial\Omega}dS_{\alpha}\sqrt{-g}A^{\alpha}+\int_{\Omega}d^{4}x\sqrt{-g}\,\widetilde{Q}^{\alpha}A_{\alpha}\ .
\end{align}
\subsection{Commutators of covariant derivatives and variations}\label{F_4}
We perform useful commutators between LC and STG covariant derivatives with variations with respect to metric and STG connection, i.e., 
\begin{align}
[D_{\alpha},\delta_{g}]A_{\beta}\bigl(g,\Gamma\bigr)&=A_{\lambda}\bigl(g,\Gamma\bigr)\delta_{g}\hat{\Gamma}^{\lambda}_{\phantom{\lambda}\alpha\beta}\ ,\\
[D_{\alpha},\delta_{g}]A^{\beta}\bigl(g,\Gamma\bigr)&=-A^{\lambda}\bigl(g,\Gamma\bigr)\delta_{g}\hat{\Gamma}^{\alpha}_{\phantom{\alpha}\lambda\alpha}\ ,\\
[\nabla_{\alpha},\delta_{g}]A_{\beta}\bigl(g,\Gamma\bigr)&=[\nabla_{\alpha},\delta_{g}]A^{\beta}\bigl(g,\Gamma\bigr)\ ,\\
[\nabla_{\alpha},\delta_{g}]B_{\mu\nu}\bigl(g,\Gamma\bigr)&=[\nabla_{\alpha},\delta_{g}]B^{\mu\nu}\bigl(g,\Gamma\bigr)=0\ ,\\
[\nabla^{\alpha},\delta_{g}]B_{\mu\nu}\bigl(g,\Gamma\bigr)&=g^{\alpha\beta}[\nabla_{\beta},\delta_{g}]B_{\mu\nu}-\nabla_{\beta}B_{\mu\nu}\delta g^{\alpha\beta}=-\nabla_{\beta}B_{\mu\nu}\delta g^{\alpha\beta}\ ,\\
[\nabla^{\alpha},\delta_{g}]B^{\mu\nu}\bigl(g,\Gamma\bigr)&=-\nabla_{\beta}B^{\mu\nu}\delta_{g}g^{\alpha\beta}\ ,\\
[D_{\alpha},\delta_{\Gamma}]A_{\beta}\bigl(g,\Gamma\bigr)&=[D_{\alpha},\delta_{\Gamma}]A^{\beta}\bigl(g,\Gamma\bigr)=0\ ,\\
[\nabla_{\alpha},\delta_{\Gamma}]A_{\beta}\bigl(g,\Gamma\bigr)&=A_{\lambda}\bigl(g,\Gamma\bigr)\delta_{\Gamma}\Gamma^{\lambda}_{\alpha\beta}\ ,\\
[\nabla_{\alpha},\delta_{\Gamma}]A^{\beta}\bigl(g,\Gamma\bigr)&=-A^{\lambda}\bigl(g,\Gamma\bigr)\delta_{\Gamma}\Gamma^{\alpha}_{\lambda\alpha}\ ,\\
[\nabla_{\alpha},\delta_{\Gamma}]B_{\mu\nu}\bigl(g,\Gamma\bigr)&=B_{\lambda\nu}\delta_{\Gamma}\Gamma^{\lambda}_{\phantom{\lambda}\alpha\mu}+B_{\mu\lambda}\delta_{\Gamma}\Gamma^{\lambda}_{\phantom{\lambda}\alpha\nu}\ ,\\
[\nabla_{\alpha},\delta_{\Gamma}]B^{\mu\nu}\bigl(g,\Gamma\bigr)&=-B^{\lambda\nu}\delta_{\Gamma}\Gamma^{\mu}_{\phantom{\mu}\alpha\lambda}-B^{\lambda\mu}\delta_{\Gamma}\Gamma^{\nu}_{\phantom{\nu}\alpha\lambda}\ ,\\
[\nabla_{\alpha},\delta_{\Gamma}]B_{(\mu\nu)}\bigl(g,\Gamma\bigr)&=2B_{\lambda(\mu}\delta_{\Gamma}\Gamma^{\lambda}_{\phantom{\lambda}\nu)\alpha}\ ,\\
[\nabla_{\alpha},\delta_{\Gamma}]B^{(\mu\nu)}\bigl(g,\Gamma\bigr)&=-2B^{\lambda(\mu}\delta_{\Gamma}\Gamma^{\nu)}_{\phantom{\nu}\alpha\lambda}\ ,\\
[\nabla^{\alpha},\delta_{\Gamma}]B_{\mu\nu}\bigl(g,\Gamma\bigr)&=g^{\alpha\beta}[\nabla_{\beta},\delta_{\Gamma}]B_{\mu\nu}\bigl(g,\Gamma\bigr)\ ,\\
[\nabla^{\alpha},\delta_{\Gamma}]B^{\mu\nu}\bigl(g,\Gamma\bigr)&=g^{\alpha\beta}[\nabla_{\beta},\delta_{\Gamma}]B^{\mu\nu}\bigl(g,\Gamma\bigr)\ .
\end{align}
\subsection{Variations}\label{F_5}
We perform the variations with respect to the metric tensor $g_{\mu\nu}$, namely, 
\begin{align}
\delta g_{\mu\nu}=&-g_{\mu\alpha}g_{\nu\beta}\delta g^{\alpha\beta}\ ,\\
\partial_{\lambda}g_{\mu\nu}\delta g^{\mu\nu}=\,&\partial_{\lambda}g^{\mu\nu}\delta g_{\mu\nu}\ ,\\
\delta_{g}Q^{\alpha}=\,&\Bigl(\delta^{\alpha}_{\mu}Q_{\nu}-Q^{\alpha}_{\phantom{\alpha}\mu\nu}\Bigr)\delta g^{\mu\nu}-g_{\mu\nu}\nabla^{\alpha}\delta g^{\mu\nu}\ ,\label{8.9}\\
\delta_{g}\widetilde{Q}^{\alpha}=&-\nabla_{\rho}\delta g^{\rho\alpha}\ ,\\
\delta_{g}Q_{\alpha}=\,&-Q_{\alpha\mu\nu}\delta g^{\mu\nu}-g_{\mu\nu}\nabla_{\alpha}\delta g^{\mu\nu}\ ,\\
\delta \sqrt{-g}=&-\frac{1}{2}\sqrt{-g}g_{\mu\nu}\delta g^{\mu\nu}\ ,
\end{align}
\begin{align}
\delta_{g}\Bigl(Q_{\alpha\beta\gamma}Q^{\alpha\beta\gamma}\Bigr)=&2Q^{\alpha}
_{\phantom{\alpha}\mu\nu}\delta_{g}Q_{\alpha}
^{\phantom{\alpha}\mu\nu}+\Big(Q_{\mu}^{\phantom{\mu}\beta\gamma}Q_{\nu\beta\gamma}-2Q_{\alpha\mu\gamma}Q^{\alpha\phantom{\nu}\gamma}_{\phantom{\alpha}\nu}\Bigr)\delta g^{\mu\nu}\ ,\\
\delta_{g}\Bigl(Q_{\alpha\beta\gamma}Q^{\beta\alpha\gamma}\Bigr)=&2Q_{(\mu\phantom{\alpha}\nu)}^{\phantom{\mu}\alpha}\delta_{g}Q_{\alpha}
^{\phantom{\alpha}\mu\nu}-Q_{\alpha\beta\mu}Q^{\beta\alpha}_{\phantom{\beta\alpha}\nu}\delta g^{\mu\nu}\ ,\\
\delta_{g}\Bigl(Q_{\alpha}Q^{\alpha}\Bigr)=&2g_{\mu\nu}Q^{\alpha}\delta_{g}Q_{\alpha}
^{\phantom{\alpha}\mu\nu}+\Bigl(Q_{\mu}Q_{\nu}-2Q^{\alpha}Q_{\alpha\mu\nu}\Bigr)\delta g^{\mu\nu}\ ,\\
\delta_{g}\Bigl(Q_{\alpha}\widetilde{Q}^{\alpha}\Bigr)=&\Bigl(\delta^{\alpha}_{(\mu}Q_{\nu)}+g_{\mu\nu}\widetilde{Q}^{\alpha}\Bigr)\delta_{g}Q_{\alpha}
^{\phantom{\alpha}\mu\nu}-\widetilde{Q}^{\alpha}Q_{\alpha\mu\nu}\delta g^{\mu\nu}\ ,\\
\delta_{g}Q=&\Bigl(P_{\mu\alpha\beta}Q_{\nu}^{\phantom{\nu}\alpha\beta}-2Q^{\alpha\beta}_{\phantom{\alpha\beta}\mu}P_{\alpha\beta\nu}\Bigr)\delta g^{\mu\nu}-2P^{\alpha}_{\phantom{\alpha}\mu\nu}\nabla_{\alpha}\delta g^{\mu\nu}\ , \label{8.10}
\end{align}
\begin{equation}
2f_{Q}\sqrt{-g}P^{\alpha}_{\phantom{\alpha}\mu\nu}\nabla_{\alpha}\delta g^{\mu\nu}=\mathcal{D}_{\alpha}\bigl(2f_{Q}\sqrt{-g}P^{\alpha}_{\phantom{\alpha}\mu\nu}\delta g^{\mu\nu}\bigr)-2\nabla_{\alpha}\bigl(\sqrt{-g}f_{Q}P^{\alpha}_{\phantom{\alpha}\mu\nu}\bigr)\delta g^{\mu\nu}\ ,
\end{equation}
\begin{equation}\label{8.11}
\delta_{g}B=\mathcal{D}_{\alpha}\Bigl[\bigl(-Q^{\alpha}_{\phantom{\alpha}\mu\nu}+\delta^{\alpha}_{\mu}Q_{\nu}\bigr)\delta g^{\mu\nu}\Bigr]+\mathcal{D}_{\mu}\nabla_{\nu}\delta g^{\mu\nu}-g_{\mu\nu}\mathcal{D}_{\alpha}\nabla^{\alpha}\delta g^{\mu\nu}+B^{\alpha}\delta\hat{\Gamma}^{\rho}_{\phantom{\rho}\alpha\rho}\ . 
\end{equation}
Now, we calculate the variation with respect to the STG connection $\Gamma^{\alpha}_{\phantom{\alpha}\mu\nu}$, i.e.,
\begin{align}
\delta_{\Gamma}R^{\alpha}_{\phantom{\alpha}\beta\mu\nu}&=2\nabla_{[\mu}\delta\Gamma^{\alpha}_{\phantom{\alpha}\nu]\beta}+T^{\gamma}_{\phantom{\gamma}\mu\nu}\delta\Gamma^{\alpha}_{\phantom{\alpha}\gamma\beta\ ,}\label{8.60}\\
\delta_{\Gamma}T^{\alpha}_{\phantom{\alpha}\mu\nu}&=2\delta\Gamma^{\alpha}_{\phantom{\alpha}[\mu\nu]}\ ,\label{8.61}\\
\delta_{\Gamma}f(Q)&=-4f_{Q}P^{\mu\nu}_{\phantom{\mu\nu}\alpha}\delta\Gamma^{\alpha}_{\phantom{\alpha}\mu\nu}\ ,\label{8.62}\\
\delta_{\Gamma}\bigl(Q^{\rho}_{\phantom{\rho}\beta\gamma}\bigr)&=-g^{\mu\rho}\bigl[\delta_{\gamma}^{\nu}g_{\beta\alpha}+\delta_{\beta}^{\nu}g_{\alpha\gamma}\bigr]\delta\Gamma^{\alpha}_{\phantom{\alpha}\mu\nu}\ ,\label{8.63}\\
\delta_{\Gamma}Q^{\rho}&=-2\delta_{\alpha}^{\nu}g^{\mu\rho}\delta\Gamma^{\alpha}_{\phantom{\alpha}\mu\nu}\ ,\label{8.64}\\\
\delta_{\Gamma}\widetilde{Q}^{\rho}&=-\bigl[\delta_{\alpha}^{\rho}g^{\mu\nu}+\delta_{\alpha}^{\mu}g^{\nu\rho}\bigr]\delta\Gamma^{\alpha}_{\phantom{\alpha}\mu\nu}\ ,\label{8.65}\\
\delta_{\Gamma}B&=\bigl[g^{\mu\nu}\mathcal{D}_{\alpha}-2\delta_{\alpha}^{\nu}\mathcal{D}^{\mu}+\delta_{\alpha}^{\mu}\mathcal{D}^{\nu}\bigr]\delta\Gamma^{\alpha}_{\phantom{\alpha}\mu\nu}\nonumber\label{8.66}\\
&=2\bigl[g^{\rho[\nu}\delta_{\alpha}^{\mu]}+g^{\mu[\nu}\delta_{\alpha}^{\rho]}\bigr]\mathcal{D}_{\rho}\delta\Gamma^{\alpha}_{\phantom{\alpha}\mu\nu}\ .
\end{align}
 \section*{Acknowledgements}
The Authors acknowledge the Istituto Nazionale di Fisica Nucleare (INFN) Sez. di Napoli,  Iniziative Specifiche QGSKY and MoonLight2, and the Istituto Nazionale di Alta Matematica (INdAM), gruppo Gruppo Nazionale di Fisica Matematica.
This paper is based upon work from COST Action CA21136: {\it Addressing observational tensions in cosmology with systematics and fundamental physics} (CosmoVerse) supported by COST (European Cooperation in Science and Technology).

\end{document}